\documentclass[11pt,a4paper]{article}
\usepackage{jheppub}
\usepackage[T1]{fontenc}
\usepackage{algorithm}
\usepackage{algorithmic}
\usepackage{hyperref}
\usepackage{graphicx}
\usepackage[font=small]{caption}
\newcommand{\diag}[3]{\raisebox{#3cm}{\includegraphics[width=#2cm]{figures/#1.eps}}}

\begin{document}
\title{Final-state QED Multipole Radiation in Antenna Parton Showers}
\author{Ronald Kleiss}
\author{and Rob Verheyen}
\affiliation{Institute for Mathematics, Astrophysics and Particle Physics, \\
Faculty of Science, Mailbox 79, Radboud University Nijmegen, \\
P.O. Box 9010, NL-6500 GL Nijmegen, The Netherlands}

\arxivnumber{1709.04485}
\abstract{We present a formalism for a fully coherent QED parton shower. The complete multipole structure of photonic radiation is incorporated in a single branching kernel. The regular on-shell $2 \rightarrow 3$ kinematic picture is kept intact by dividing the radiative phase space into sectors, allowing for a definition of the ordering variable that is similar to QCD antenna showers. A modified version of the Sudakov veto algorithm is discussed that increases performance at the cost of the introduction of weighted events. Due to the absence of a soft singularity, the formalism for photon splitting is very similar to the QCD analogon of gluon splitting. However, since no color structure is available to guide the selection of a spectator, a weighted selection procedure from all available spectators is introduced. }

\maketitle

\section{Introduction}
Phenomenology for high energy particle collisions such as those at the LHC is currently strongly reliant on the implementation of theoretical knowledge in Monte Carlo based event generators \cite{manual}. In these programs, the simulation of collisions is partitioned into several tasks, of which one is the parton shower. These programs dress hard events with soft and collinear radiation, resumming the associated leading logarithms, and preparing them for hadronization. The Standard Model allows for many types of radiation, of which QCD radiation is the most important in the context of LHC phenomenology. For that reason, QCD parton showers have been developed for decades. Traditionally, parton showers were based on the DGLAP equation \cite{dglap1,dglap2,dglap} and $1 \rightarrow 2$ kinematics. Soft coherence can then be ensured by some form of angular ordering \cite{coherence1,coherence2}. More recently, showers based on $2 \rightarrow 3$ kinematics have become increasingly popular. These showers also obey the DGLAP equation, but are automatically soflty coherent and have access to exact phase space factorization. 

QCD parton showers based on $2 \rightarrow 3$ factorization can be divided into two categories. On the one hand are the \emph{antenna showers} based on the antenna factorization scheme for fixed order calculations \cite{antenna1Map, antenna2,antenna3, antenna4, antenna5}. These showers include the ARIADNE code \cite{ariadne} which was very successful in the LEP era, the SHERPA-based ANTS shower \cite{ants}, and more recently the VINCIA shower \cite{vinciaMassless, vinciaMassive, vinciaHadron, vinciaSector}. Antenna showers treat the partons participating in a branching on the same footing. From the factorization properties of matrix elements, an antenna function is derived which serves as probability measure for branchings. These branching kernels capture both the soft and collinear divergent behaviour of the matrix element associated with the branchings. 

On the other hand are the \emph{dipole showers} based on the Catani-Seymour factorization scheme \cite{CS1,CS2}. These types of showers are currently included in the SHERPA \cite{sherpaCS} and HERWIG \cite{herwigCS} event generators. Dipole showers essentially divide the antenna function into two pieces that radiate independently. For both of these kernels, one of the participating partons acts as a recoiler, while the other parton branches. 

Finally, the recently developed DIRE \cite{dire} parton shower is a DGLAP-based shower that contains all the advantages of dipole and antenna showers. It achieves soft coherence by using modified Altarelli-Parisi kernels and uses $2 \rightarrow 3$ for exact momentum conservation. 

Other types of radiation are allowed in the Standard Model. This radiation should be interleaved with the dominant QCD radiation. Most event generators have implementations for QED radiation \cite{photos,photonSherpa,photonPythia}, which includes the emission of photons from charged fermions, and the splitting of photons into fermion-antifermion pairs. However, photon emission is not formally coherent in any of these implementation. In leading-color QCD DGLAP-style showers, coherence can be achieved by angular ordering, but this does not straightforwardly extend to QED. In antenna or dipole approaches, the branching kernel cannot be partitioned into independently radiating pieces, since there is no color structure distinguishing their post-branching states. On the other hand, YFS exponentiation \cite{yfs} is used by some event generators \cite{decays1,decays2} to add soft photon radiation to particle decays, and by others to simulate process-specific photon radiation for precision physics \cite{yfsmc,ceex,kkmc,herwiri}. This type of radiation is coherent, but collinear logarithms can only be included order-by-order, and it cannot be interleaved with QCD radiation. In this paper, we introduce a parton shower formalism which produces coherent QED radiation and which can be interleaved in a regular QCD shower. Our formalism does not nessecarily fit into either the antenna or dipole categories, but since its kinematical approach is closer to an antenna shower, and to VINCIA in particular, we refer to it as such. 

In section \ref{emissionSection} the shower formalism for photonic radiation is introduced. A comparison with leading-color QCD showers is used to explain the complications that have to be dealt with in a fully coherent photon shower. The algorithmic implementation of the shower and a method of improving performance are also explained. In section \ref{splittingSection} the shower formalism for photon splitting into charged fermion-antifermion pairs is explained. This type of radiation is very similar to gluon splitting, with the exception of the absence of a color structure to dictate a spectator parton. The shower formalism is tested in various ways in section \ref{resultsSection}, including comparison with exact matrix element calculations, the DGLAP equation and YFS radiation patterns. Finally, section \ref{conclusionSection} contains a summary and some outlook. 

\subsection{Notation and conventions}
For momenta $p_i$ and $p_j$ and masses $m_i$ and $m_j$ we will make extensive use of the notations
\begin{equation}
s_{ij} = 2p_i {\cdot}p_j \quad m_{ij}^2 = (p_i + p_j)^2 = s_{ij} + m_i^2 + m_j^2.
\end{equation} 

In a massive $2 \rightarrow 3$ branching, we denote the pre-branching system with upper case letters and the post branching-system with lower case letters.

\section{Photon emissions} \label{emissionSection}
All antenna and dipole showers function in the leading-color QCD approximation. In this context, it makes sense to partition the total gluon contribution into antennae or dipoles corresponding with differing color ordered states. Because of the leading-color approximation, the number of contributing antennae or dipoles is limited and their interference structure is automatically disentangled. 

\begin{figure}[H]
\begin{align}
\diag{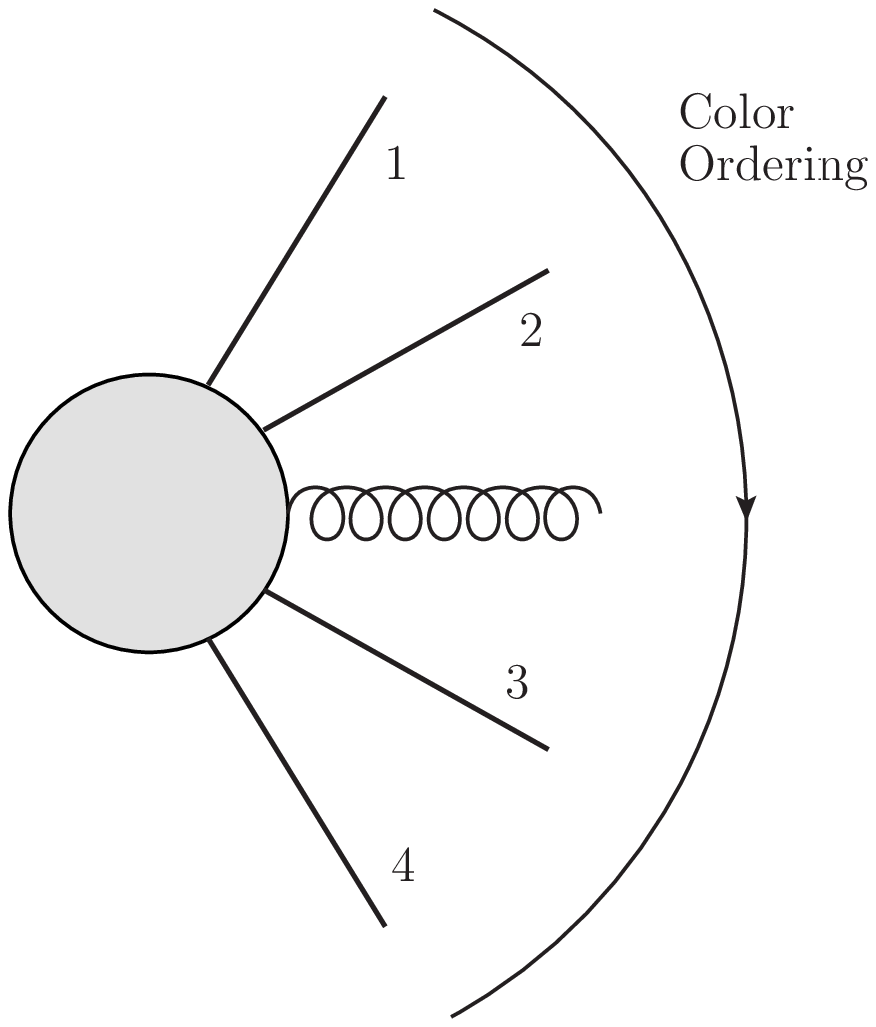}{3.2}{-1.85}	\xrightarrow[\text{collinear}]{\text{soft}} \diag{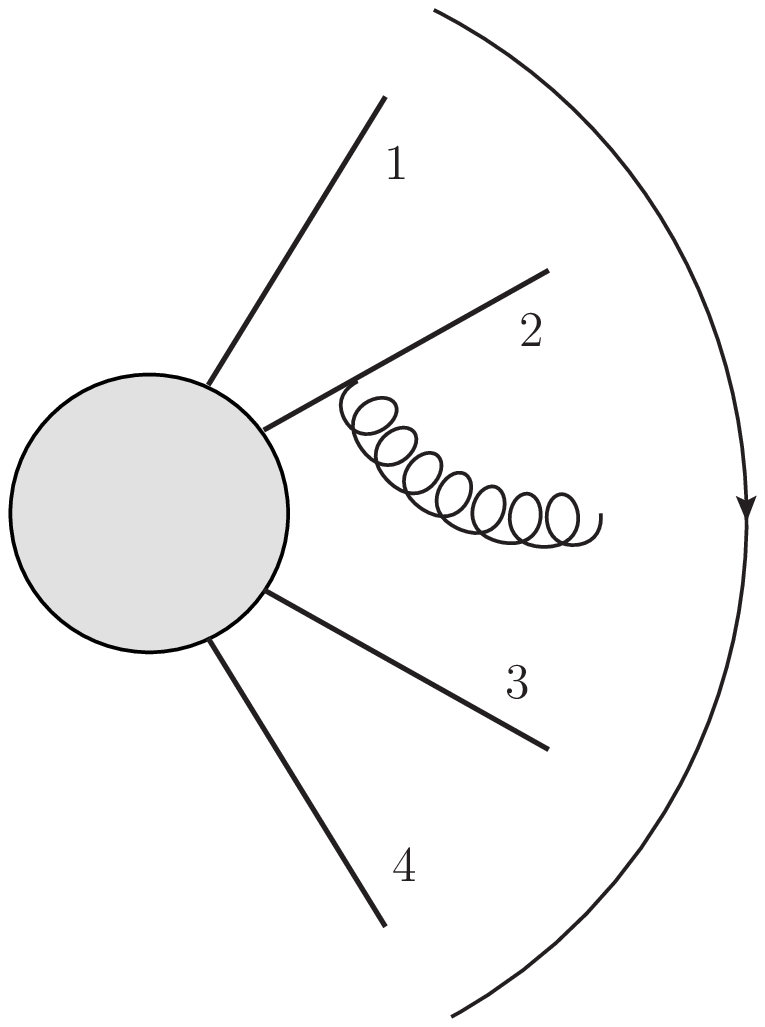}{2.9}{-1.8} + \diag{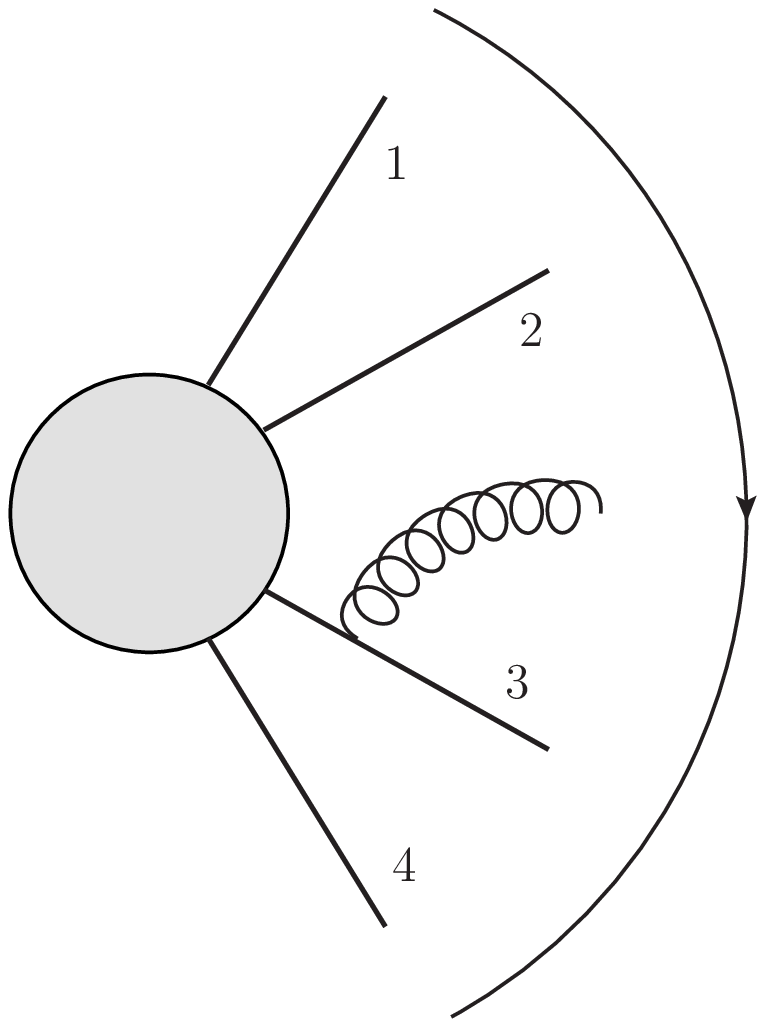}{2.9}{-1.8} + \mathcal{O}\left( \frac{1}{N_c}\right) \nonumber 
\end{align}
\caption{Factorization of soft or collinear gluon emission in leading-color QCD.}
\end{figure}

In contrast, for photon emissions, there is no color structure or leading color approximation. This means that every pair of charged (anti)fermions contributes equally, and there is no way to divide the kernel into several disconnected pieces. As a consequence, every charged fermion in principle participates in the emission of a photon. 

\begin{figure}[H]
\begin{align}
\diag{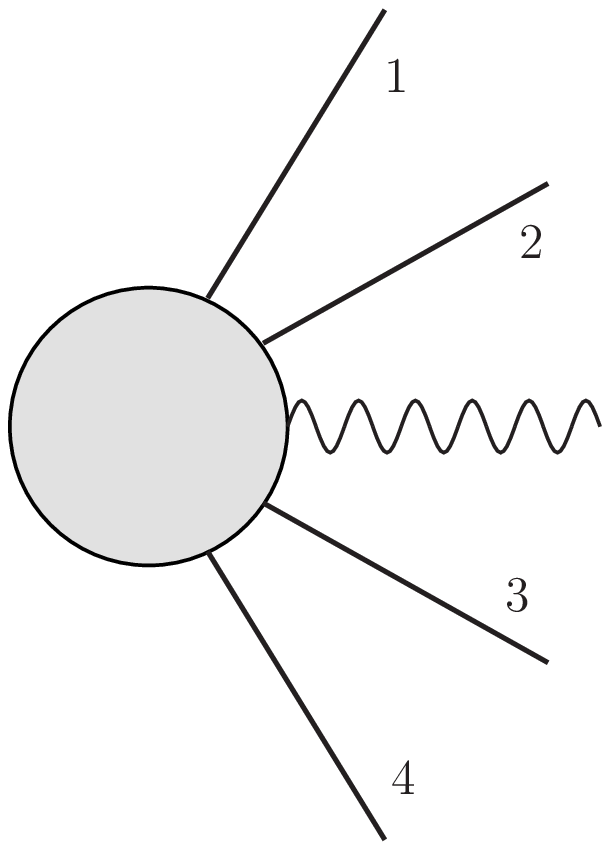}{2.5}{-1.4} \xrightarrow[\text{collinear}]{\text{soft}} \diag{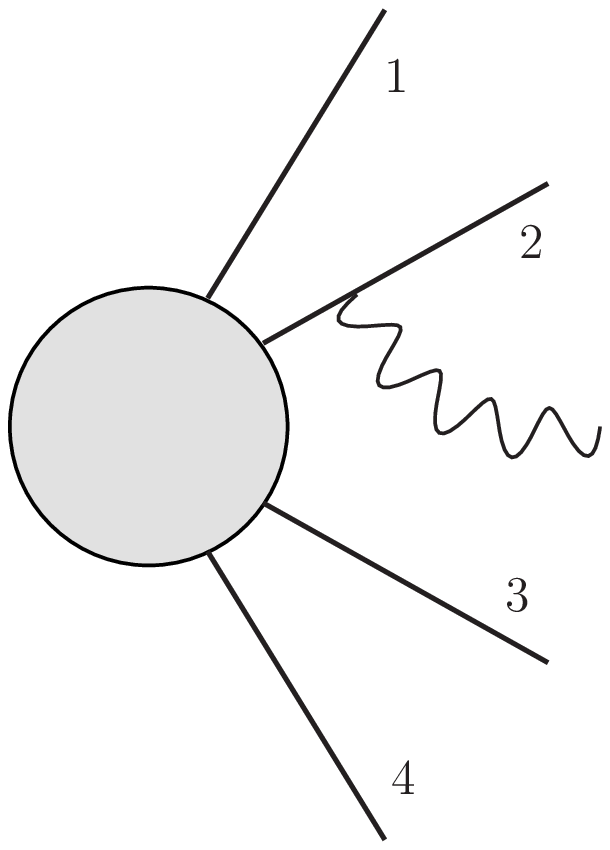}{2.5}{-1.4} + \diag{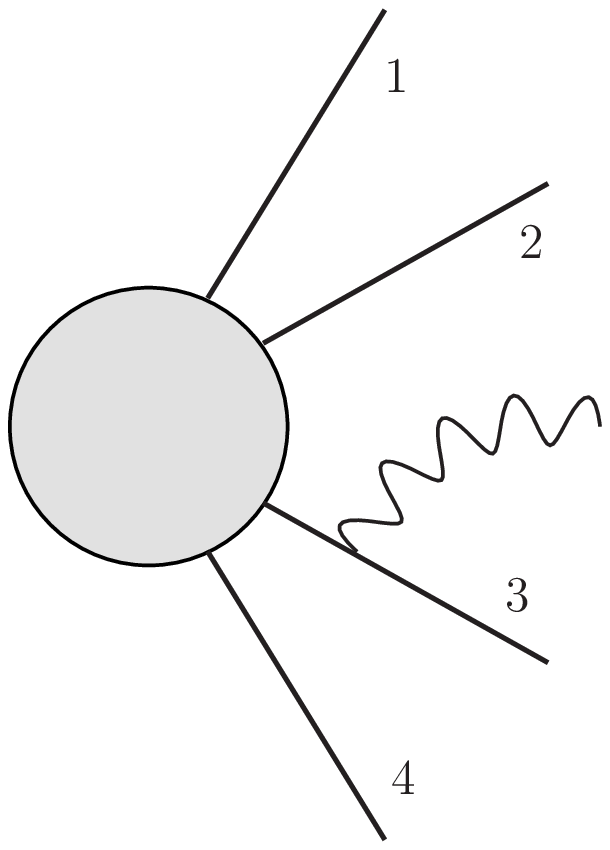}{2.5}{-1.4} + \diag{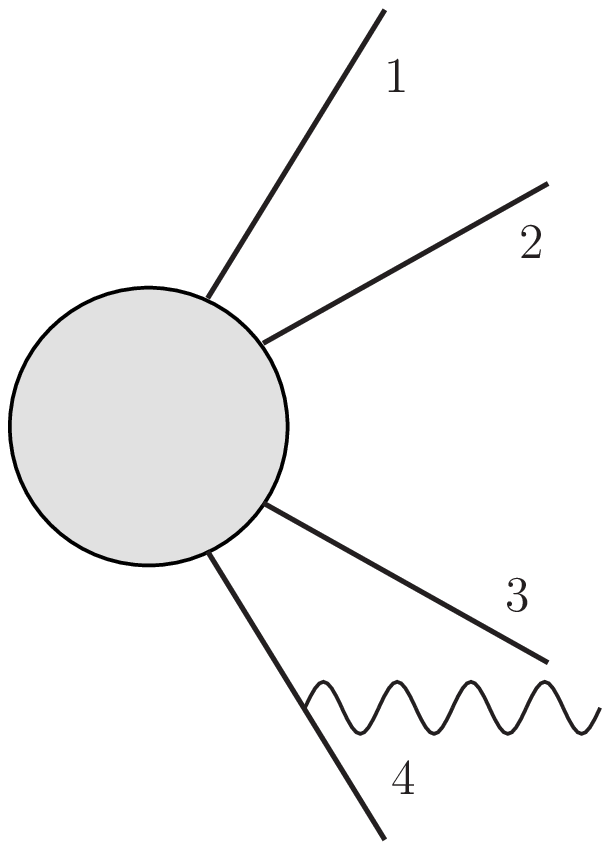}{2.5}{-1.4} + \diag{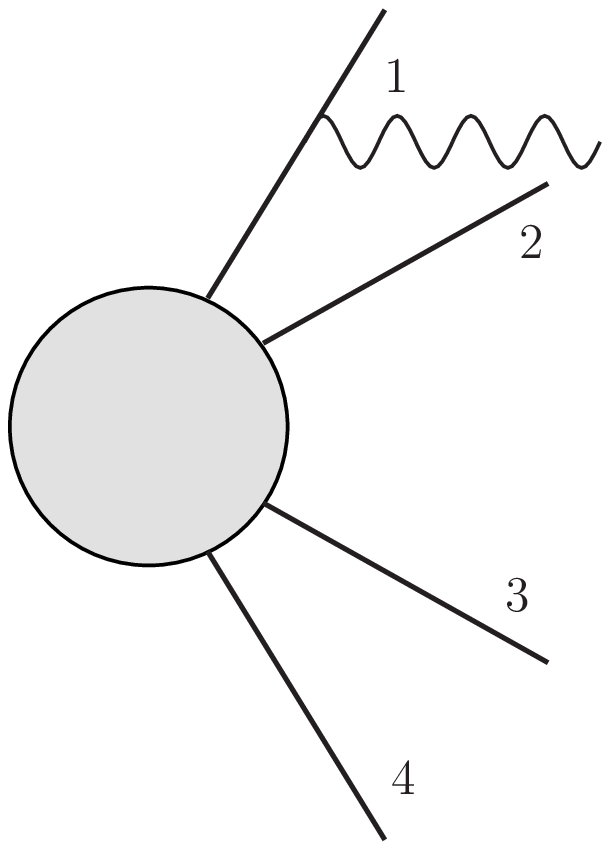}{2.5}{-1.4}\nonumber
\end{align}
\caption{Factorization of soft or collinear photon emission in QED.}
\end{figure}

The implementation of such a procedure in a parton shower formalism would be problematic, especially if it is to be interleaved with a regular QCD shower. For that reason, we will employ an approach similar to the sector shower detailed in \cite{vinciaSector} to cast photon emissions in an antenna shower-like procedure. Below, we first derive the emission kernel, and then proceed to detail the shower implementation by the definition of the ordering variable and the kinematics.

\subsection{Emission kernel}
Let $|M_1(\{p\},k)|^2$ be a squared amplitude of a set of $n$ final-state charged fermions and antifermions with momenta $p_i$ and charges $Q_i$, and a photon with momentum $k$. The factorization properties of such amplitudes are well-known \cite{photonFactorization}. In the soft limit $k \rightarrow 0$, the matrix element factorizes into a photonic piece which is usually called the \emph{eikonal} factor, and a lower order matrix element that does not include the photon: 
\begin{equation} \label{softQED}
|M_1(\{p\},k)|^2 = -4\pi \alpha \sum_{a,b=1}^n 2 Q_a Q_b \frac{s_{ab}}{s_{ak}s_{bk}} |M_0(\{p\})|^2
\end{equation}
where $\alpha$ is the fine-structure constant. The soft factorization of the eikonal factor is a general property of QED matrix elements and serves as one of the bases of the shower. We further note that the momentum of every charged particle appears in the eikonal factor, but they remain unchanged since the photon is soft. This property must be reflected in the parton shower. Factorization of the matrix element also occurs in the the quasi-collinear limit, which is the massive generalization of the collinear limit \cite{quasiCollinear,antenna5}. For momentum $p_a$ it is defined as 
\begin{align} \label{quasiCollinear}
k \rightarrow (1-z) \widetilde{p}_a &\qquad p_a \rightarrow z\widetilde{p}_a \nonumber \\
\intertext{while imposing}
s_{ak}, \, m_a, \,\widetilde{m}_a \rightarrow 0& \mbox{ at fixed ratios }\frac{m_a^2}{s_{ak}}, \, \frac{\widetilde{m}_a^2}{s_{ak}}.
\end{align}
Compared with the massless collinear limit, the main difference is that $m_a^2$ and $\widetilde{m}_a^2$ should be kept of the same order while $\widetilde{m}_a^2$ tends to zero. In this limit the matrix element factorizes into a quasi-collinear piece involving only $k$ and $p_a$, and the remaining photon-less matrix element with $\widetilde{p}_a$: 
\begin{equation} \label{collinearQED}
|M_1(\{p\},k)|^2 = 4\pi \alpha \, Q_a^2  \frac{2}{s_{ak}} P(z, m_a^2,s_{ak}) |M_0(p_1,..,\widetilde{p}_a,..,p_n)|^2
\end{equation}
where
\begin{equation}
P(z, m_a^2,p_a{\cdot}k) = \frac{1+(1-z)^2}{z} - \frac{2m_a^2}{s_{ak}}
\end{equation}
is the generalized Altarelli-Parisi splitting kernel. Note that in this limit, $p_a$ is the only momentum participating in the emission. Again, the parton shower should reflect this property. Similar factorization theorems apply in QCD. However, in the leading-color approximation, only terms involving partons which are adjacent in color space survive. The radiative factors of soft and collinear gluon emission inbetween two partons $a$ and $b$ can be combined into an antenna function
\begin{equation} \label{aQCD}
a_{e}^{\mbox{\tiny{QCD}}}(p_a,k,p_b) = C \bigg[ 2\frac{s_{ab}}{s_{ak}s_{bk}} - 2\frac{m_a^2}{s_{ak}^2} - 2\frac{m_b^2}{s_{bk}^2}+ \frac{1}{s_{ak} + s_{bk} + s_{ab}} \left( \frac{s_{ak}}{s_{bk}} + \frac{s_{bk}}{s_{ak}} \right)\bigg]
\end{equation}
 where $k$ is the gluon momentum and $C$ is a color factor. This function captures the soft and the collinear factorization properties of a partial amplitude in the leading-color limit. The total gluon emission amplitude can then be approximated by summing over such an antenna function for every possible pair of partons $\{ab\}$. The functional form of $a_{e}^{QCD}$ can be computed from matrix elements \cite{antenna4,antenna5} as 
\begin{equation}
a_{e}^{\mbox{\tiny{QCD}}}(p_a,k,p_b) = \frac{|M(X \rightarrow \bar{q}gq)|^2}{4\pi \alpha_{s} |M(X \rightarrow \bar{q}q)|^2}
\end{equation}
where $X$ is some decaying particle such as a $Z$ or $\gamma^*$. Depending on $X$, the computation yields different non-singular terms as part of the antenna function. These terms serve as tuning parameters in the parton shower. 

There is no leading-color limit in QED, and none of the eikonal factors of eq.~\eqref{softQED} is of lesser importance. It is therefore not feasible to describe the emission of a photon as a sum of antennae without resorting to approximations. We instead capture all factorization properties in a single emission kernel. Similar to QCD, the singular structure for photon emissions is extracted from a matrix element calculation. The total emission kernel up to nonsingular terms is then just the antenna function summed over all charged pairs. 
\begin{equation} \label{aQED} 
a_{e}^{\mbox{\tiny{QED}}}(\{p\},k) = \sum_{\{ab\}} -2Q_a Q_b \bigg[ 2\frac{s_{ab}}{s_{ak}s_{bk}} - 2\frac{m_a^2}{s_{ak}^2} - 2\frac{m_b^2}{s_{bk}^2}+ \frac{1}{s_{ak} + s_{bk} + s_{ab}} \left( \frac{s_{ak}}{s_{bk}} + \frac{s_{bk}}{s_{ak}} \right) \bigg],
\end{equation}
where we note that additional nonsingular terms may be included depending on the process used to calculate the individual terms. While in eq.~\eqref{softQED} the indices of the sum run over all charged fermions, here the sum runs over all \emph{pairs} $\{ab\}$. The $a=b$ terms in eq.~\eqref{softQED} lead to the mass-dependent terms in eq.~\eqref{aQED} which are redistributed amongst the pairwise sum using charge conservation. 

\subsection{Ordering variable and phase space}
We now cast the result in the standard antenna shower formalism \cite{vinciaMassless,ariadne,ants} including fermion mass effects. At its core lies a $2 \rightarrow 3$ branching step which yields on-shell massive momenta and obeys momentum conservation. The probability for emissions are proportional to the emission kernel of eq.~\eqref{aQED}, weighted with a Sudakov factor which resums the leading logarithms associated with the soft and collinear singularities. To regulate subsequent branchings, an ordering variable is defined which acts as a resolution scale \cite{ordering}. As the shower runs, the ordering variable decreases and branchings occur, resolving increasingly soft and collinear radiation. At some point, the ordering variable hits a cutoff value, terminating the parton shower. Here, we first briefly describe the antenna shower formalism for leading-color gluon emissions, before continuing with photon emissions.

We denote the participating momenta as 
\begin{equation}
p_A + p_B \rightarrow p_a + p_b + p_k.
\end{equation}
where $p_A^2 = p_a^2 = m_a^2$ and $p_B^2 = p_b^2 = m_b^2$. The momentum $p_k$ is the emitted gauge boson and has $p_k^2 = 0$. 
\subsubsection{Leading-color QCD}
For leading-color gluon emissions a sensible choice for the ordering variable is the transverse momentum with respect to the pre-branching system. It is defined as
\begin{equation} \label{tEmission}
t = (p^2_{\perp})_{ab} = 4\frac{s_{ak} s_{bk}}{m^2_{abk}}
\end{equation}
where the factor $4$ is included to ensure $t \leq m_{AB}^2$. This ordering variable is used in \cite{ariadne,ants} up to a factor, and is is one of the available options in \cite{vinciaMassive}. Most importantly, all singular behaviour is contained in the phase space region where $t \rightarrow 0$. A cutoff on the scale therefore effectively regularizes the singularities, preventing situations where the emission probability diverges. 

Next, we also have to factorize the $(n+1)$-body phase space $d\Phi_{n+1}$ to completely separate the radiative piece of a cross section. To this end the antenna factorization
\begin{equation} \label{antennaPhaseSpace}
d\Phi_{n+1} = d\Phi^{\mbox{\tiny{ant}}}_{ab} d\Phi_n
\end{equation}
is used, where
\begin{align}
d\Phi^{\mbox{\tiny{ant}}}_{ab} &= \frac{1}{16 \pi^2}\lambda^{-\frac{1}{2}}(m_{uv}^2,m_u^2,m_v^2)\,ds_{ab} \,ds_{ac} \,ds_{bc} \frac{d\phi}{2\pi} \nonumber \\
&\times \delta\left(m_{abc}^2-s_{ab}-s_{ac} - s_{bc} - m_a^2 - m_b^2 - m_c^2\right) \theta\left(G_{abc}>0\right).
\end{align}
where $G_{ab} = s_{ab}s_{bc}s_{ac}-s_{ab}^2m_c^2 - s_{ac}^2m_{b}^2 - s_{bc}^2m_{a}^2 + 4m_{a}^2m_{b}^2m_{c}^2$ is the three-body Gram determinant. A brief derivation of this factorization is sketched in Appendix \ref{PSF}. We can now write down the \emph{shower approximation} for a gluonic color ordered matrix element as
\begin{equation}
|M_{n+1}(..,p_a,k,p_b,..)|^2 d\Phi_{n+1} \approx 4\pi \alpha_s a_e^{\mbox{\tiny QCD}}(p_a,k,p_b) \, d\Phi^{\mbox{\tiny{ant}}}_{ab} \, |M_{n}(..,p_A, p_B,..)|^2 d\Phi_n.
\end{equation}
The crucial point here is that this equality is exact in the singular regions of phase space where the shower aims to correctly reproduce the matrix element, but merely approximate elsewhere. The antenna phase space is then transformed to include the ordering variable $t$. To this end, an auxilliary variable 
\begin{equation}
z = \frac{s_{ak}}{s_{ak} + s_{bk}}
\end{equation}
is introduced. If all particles are massless, the boundaries of $z$ are given by
\begin{equation}
\frac{1}{2}\left(1 - \sqrt{1 - \frac{t}{m^2_{AB}}} \right) = z_- \leq z \leq z_+ = \frac{1}{2}\left(1 + \sqrt{1 - \frac{t}{m^2_{AB}}} \right).
\end{equation}
However, for massive particles, these expressions are more complicated and are most conveniently given by the positivity of the Gram determinant. We note that the functional form of $z$ has no physical effect and is only chosen for convenience. This particular choice provides a clear physical picture in the sense that for a given value of $t$, $z \rightarrow 0$ and $z \rightarrow 1$ correspond with the collinear phase space regions, while the soft region lies inbetween. Transforming the phase space to these variables introduces a Jacobian $|J| = \frac{m^2_{AB}}{8z(1-z)}$, leading to
\begin{align} \label{emitPS}
d\Phi^{\mbox{\tiny{ant}}}_{ab} = \frac{1}{128 \pi^2} dt \frac{dz}{z(1-z)} \frac{d\phi}{2\pi} \frac{m^2_{AB}}{\lambda^{\frac{1}{2}}(m_{AB}^2,m_a^2,m_b^2)} \theta(m_{AB}^2-s_{ak}-s_{bk} - m_a^2 - m_b^2) \theta(G_{abk}>0).
\end{align}
The shower proceeds by generating the shower variables $t$, $z$ and $\phi$ and selecting a color-adjacent parton pair $ab$. The pre-branching system is then transformed to the post-branching system using the generated variables and the kinematics map which we detail in section \ref{emissionKinematicsSection}. Finally, through the Sudakov veto algorithm \cite{Us,negativeVeto1,negativeVeto3}, the event passes a rejection step, giving it a probability
\begin{equation} \label{QCDprob}
S_e^{\mbox{\tiny{QCD}}}(p_a, k, p_b; u) = 4\pi \alpha_s a_{e}^{\mbox{\tiny{QCD}}}(p_a,k,p_b) \theta\left(u - t \right)\exp\left(-\sum_{\{ij\}} 4\pi \alpha_s \int_t^u d\Phi_{ij}^{\mbox{\tiny{ant}}} a_{e}^{\mbox{\tiny{QCD}}}(p_i,k,p_j)\right)
\end{equation} 
to be accepted. Here, $u$ is the upper boundary of the evolution variable.

\subsubsection{QED}
From the parton shower perspective the main difference between leading-color gluon radiation and photon radiation is the separation of the emission probability for different color structures. These probabilities can be partitioned in antennae as in eq.~\eqref{QCDprob} because they correspond with different color orderings, and therefore different final states. In QED, there is no color structure and a partitioning is not possible without discarding some of the contributions to eq.~\eqref{softQED}. However, to maintain the parton shower picture and to allow for interleaving with QCD radiation, we would prefer to use the antenna phase space factorization of eq.~\eqref{antennaPhaseSpace}, even though every charged fermion should participate in the emission. Our strategy will therefore be to divide phase space into \emph{sectors}, similar to \cite{vinciaSector}. There, this approach is instead used to simplify the matching and merging procedure of the VINCIA shower. 

The sector method is most easily understood by considering the QED sector shower approximation
\begin{equation} \label{QEDshowerApprox}
|M_{n+1}(\{p\},k)|^2 \approx \sum_{\{ab\}} 4\pi \alpha \, \Theta\left((p_\perp^2)_{ab}\right)a_e^{\mbox{\tiny QED}}((\{p\},k)_{ab})|M_{n}(\{p\})|^2 ,
\end{equation}
where
\begin{equation}
\Theta\left((p_\perp^2)_{ab}\right) =
\left\{
	\begin{array}{ll}
		1  & \mbox{if } \forall \{ij\} \,\, (p^2_{\perp})_{ab} \leq (p^2_{\perp})_{ij}\\
		0 & \mbox{else}
	\end{array}
\right.
\end{equation}
and we denote the momenta of the branching process by $\{p\} \rightarrow (\{p\},k)_{ab}$ if fermions $a$ and $b$ participate. At every phase space point, only a single term of eq.~\eqref{QEDshowerApprox} is active. In terms of the parton shower, the photon is emitted by the charged pair it has the lowest transverse momentum with. This ensures that, as long as the $2 \rightarrow 3$ kinematics are infrared and collinear safe, the emission kernel never encounters a divergence. The corresponding ordering variable is
\begin{equation} \label{QEDord}
t = \min\left( (p^2_{\perp})_{ab} \right).
\end{equation}
This ordering variable has the required property of ensuring that all soft and collinear regions are contained in the limit $t \rightarrow 0$, while still allowing for regular antenna shower kinematics. It can be implemented with an additional step in the Sudakov veto algorithm, which we discuss in section \ref{svaSection}. It will produce emissions distributed according to
\begin{align} \label{QEDprob}
S_e^{\mbox{\tiny{QED}}}(\{p\},k; u) &= \sum_{\{ab\}} 4\pi \alpha \, \Theta((p^2_{\perp})_{ab}) \theta(u - t) \, a_e^{\mbox{\tiny QED}}((\{p\},k)_{ab}) \nonumber \\
&\times \exp\left(-\sum_{\{a'b'\}} \int^u_t 4\pi \alpha \, d\Phi^{\mbox{\tiny{ant}}}_{a'b'} \Theta((p^2_{\perp})_{a'b'}) a_e^{\mbox{\tiny QED}}((\{p\},k)_{a'b'}) \right).
\end{align}
We emphasize that eq.~\eqref{QEDshowerApprox} contains the correct soft and collinear behaviour at the current scale of emission. After the shower is continued, more charged particles may be produced through photon splitting as described in section \ref{splittingSection}, or QCD radiation. In the leading logarithmic approximation, a high scale photon is blind to these lower scale effects. A similar situation appears in QCD gluon emissions, although the dynamic effects of low scale branchings are partly hidden behind the leading color approximation. Low scale branchings can still kinematically influence high scale branchings, but their effect is deemed negligible due to the ordering condition.

\subsection{Kinematics} \label{emissionKinematicsSection}
Here, we describe how the post-branching momenta are constructed from the pre-branching momenta and the shower variables. To agree with the factorization properties eq.~\eqref{collinearQED} and eq.~\eqref{softQED}, this mapping must obey the following rules:
\begin{enumerate}
\item Soft safety: For $k\rightarrow 0$, $p_a = p_A$ and $p_b = p_B$
\item Collinear safety: For $k \parallel p_a$, $p_b = p_B$ if $p_A$ is massless + equivalent for $a \leftrightarrow b$.
\end{enumerate}
Additionally, $p_a$ and $p_b$ should be treated on equal footing in the antenna picture. We use the massive generalization of the Kosower map \cite{antenna1Map} which is also used by VINCIA \cite{vinciaMassive}. The pre-branching and post-branching momenta are related by
\begin{align} \label{emissionMap}
p_A &= x_a p_a + r p_k + x_b p_b \nonumber \\
p_B &= (1-x_a) p_a + (1-r) p_k + (1-x_b)p_c. 
\end{align}
The parameters $x_a$ and $x_b$ can be fixed by setting $p_A^2 = m_a^2$ and $p_B^2 = m_b^2$, leading to
\begin{align}
x_a = \frac{1}{2}\frac{1}{4G_{abk} + m^2_{AB}(s_{ab}^2 - 4m_a^2m_b^2)} &\bigg[\Sigma^2 \left(s_{ab}^2 - 4m_a^2m_b^2 + 4\Delta_{ak}\right) + 8r\left(G_{abk} - m_{AB}^2\Delta_{ak} \right) \nonumber \\
&+ V\left(2m_{AB}(m_a - m_b) + m_{AB}^2 - m_a^2 + m_b^2 - s_{ak}\right) \bigg] \nonumber \\
x_b = \frac{1}{2}\frac{1}{4G_{abk} + m^2_{AB}(s_{ab}^2 - 4m_a^2m_b^2)} &\bigg[\Sigma^2 \left(s_{ab}^2 - 4m_a^2m_b^2 + 4\Delta_{bk}\right) + 8r\left(G_{abk} - m_{AB}^2\Delta_{bk} \right) \nonumber \\
&- V\left(2m_{AB}(m_b - m_a) + m_{AB}^2 - m_b^2 + m_a^2 - s_{ak}\right) \bigg]
\end{align}
where we defined
\begin{align}
\Sigma^2 &= m_{AB}^2 + m_a^2 - m_b^2 \nonumber \\
V^2 &= 16 G_{abk}\left(m^2_{AB} r(1-r) - (1-r)m_a^2 - rm_b^2 \right) + \left(s_{ab}^2 - 4m_a^2m_b^2\right) \left(s_{AB}^2 - 4m_a^2m_b^2\right) \nonumber \\
\Delta_{ak} &= \frac{1}{4} \left(s_{ab}s_{bk} - 2s_{ak}m_b^2 \right) \qquad \Delta_{bk} = \frac{1}{4} \left(s_{ab}s_{ak} - 2s_{bk}m_a^2 \right).
\end{align}
Next, a choice for the functional form of the parameter $r$ has to be made in such a way that the infrared and collinear safety conditions are satisfied. We follow VINCIA and use
\begin{equation}
r = \frac{1}{2m_{AB}^2}\left( \Sigma^2 + \frac{s_{bk} - s_{ak}}{s_{ak} + s_{bk}} \sqrt{s^2_{AB} - 4m_a^2 m_b^2}  \right).
\end{equation}
This choice has the property that interchanging $a \leftrightarrow b$ corresponds with $r \rightarrow 1 -r $, and it reduces to the massless Kosower map where 
\begin{equation}
r_{\mbox{\tiny{massless}}} = \frac{s_{ak}}{s_{ak} + s_{bk}}
\end{equation}
The parton shower requires the inverse of the map given by eq.~\eqref{emissionMap}. Since the shower variables and $m_{AB}^2$ fix the overal azimuthal angle of the system and the invariants $s_{ak}$, $s_{bk}$ and $s_{ab}$, which in turn fix the energies and relative angles between the post-branching momenta, the remaining choice for $r$ appears only in the angle between the pre-braching and post-branching systems, at which point the kinematics are completely fixed. One might for instance compute the angle between $p_A$ and $p_a$ as
\begin{equation} 
\cos\psi = \frac{E_A E_a - x_a m_a^2 - r s_{ak} - x_b s_{ab}}{|\vec{p}_A| |\vec{p}_a|}
\end{equation}
which has an explicit dependence on $r$. The momenta are constructed in the center of mass frame of the pre-branching system using the shower variables and this angle, and are then boosted back to the original frame. For more details on the kinematic map for emissions, we refer to VINCIA \cite{vinciaMassive}.

\subsection{Sudakov veto algorithm} \label{svaSection}
The Sudakov veto algorithm is a very important component of parton shower programs \cite{Us,negativeVeto1,negativeVeto2}. It can produce samples from probability distributions like eq.~\eqref{QCDprob} and eq.~\eqref{QEDprob} without evaluation of the integral in the exponent or explicit inversion of the cumulative density. Here, we present the veto algorithm specialized to produce the density of eq.~\eqref{QEDprob}. We refer to \cite{Us} for a more general explanation.

We first introduce some definitions to shorten the notation. As was shown in appendix \ref{PSF}, the boundaries of the emission phase space are given by the positivity of the Gram determinant $G_{abk}$. In terms of the shower variables, this tanslates to $p^2_{\perp}$-dependent boundaries on $z$. The Sudakov veto algorithm makes use of overestimates of these boundaries which are valid for any value of $t$. We use
\begin{equation}
z_{AB \pm} = \frac{1}{2}\left(1 \pm \sqrt{1 - \frac{t_{\mbox{\tiny{cut}}}}{m_{AB}^2}} \, \right)
\end{equation}
where $t_{\mbox{\tiny{cut}}}$ is the lower cutoff on the evolution variable, regulating the infrared divergences. These boundaries are based on the phase space for massless particles, which contains the massive phase space. Furthermore, we define 
\begin{equation} \label{zDef}
Z_{AB} = \int_{z_{AB-}}^{z_{AB+}} \frac{1}{z(1-z)}
\end{equation}
and channel-specific weights 
\begin{equation}
w_{AB} = \frac{1}{32 \pi} Z_{AB} \, m_{AB}^2 \, \lambda^{-\frac{1}{2}}(m_{AB}^2, m_a^2, m_b^2) \qquad W = \sum_{\{IJ\}}w_{IJ}.
\end{equation}
which contain most of the phase space factors of eq.~\eqref{emitPS}. Finally, we need an overestimate function $g(t)$ for the emission kernel, further discussed in section \ref{overestimateSection}, and independent and identically distributed random numbers $\rho_i$. The Sudakov veto algorithm for photon emissions is given in algorithm~\ref{SVAemission}. From the $p_{\perp}^2$ veto step it is immediately clear that the ordering variable is $t = \min\left( (p^2_{\perp})_{ab} \right)$ and that it is not possible for the photon to be too collinear with any of the other charged fermions. We also note that this algorithm is quite dissimilar from the standard competition algorithms mentioned in \cite{negativeVeto1,negativeVeto2}. However, algorithm~\ref{SVAemission} is essentially a competition algorithm as well, but one that facilitates competition between sectors instead of between branching kernels. It was shown in \cite{Us} that competition can be handled in different ways, and algorithm~\ref{SVAemission} is just a particularly suitable version chosen because in this case all competing channels have the same emission kernel.

\begin{algorithm} 
\caption{The Sudakov veto algorithm for photon emissions}
\label{SVAemission}
\begin{algorithmic}
\STATE $t \leftarrow u$
\LOOP
\STATE Choose a pair $\{AB\}$ with probability $w_{AB}/W$
\STATE $t \leftarrow$  solution of $\rho_1 = \exp\left(-\int_{t'}^t d\tau \, W \alpha g(\tau) \right)$ for $t'$
\STATE $z \leftarrow \left(1 + \exp\left(Z_{AB}(\frac{1}{2} - \rho_2\right)\right)^{-1}$ 
\STATE $\phi \leftarrow 2\pi \rho_3$
\STATE Compute $s_{ak}$, $s_{bk}$, and $s_{ab}$ from $t$, $z$, and $m_{AB}^2$
\IF {$G_{abk} > 0$ and $s_{ab} > 0$}
\STATE Construct the $p_a$, $p_b$ and $k$ from $p_A$, $p_B$, $t$, $z$ and $\phi$
\IF {t = ($p^2_{\perp})_{ab}$ is the smallest $p^2_{\perp}$}
\IF {$\rho_4 < a_e^{\mbox{\tiny QED}}((\{p\},k)_{ab}) / g(t)$}
\RETURN $(\{p\},k)_{ab}$
\ENDIF
\ENDIF
\ENDIF
\ENDLOOP
\end{algorithmic}
\end{algorithm}

We check if this algorithm produces the density given by eq.~\eqref{QEDprob} by writing out the probabilities step-by-step. For the sake of readability, we leave out some details such as the dependence of the post-branching momenta on the generated shower variables. 
\begin{align} \label{emissionVeto}
S_e^{\mbox{\tiny{QED}}}(\{p\},k; u) &= \frac{1}{W} \sum_{{ab}} w_{AB} \int_0^u dt \, W \alpha g(t) \exp\left( -\int^u_t d\tau W \alpha g(\tau)\right) \nonumber \\
&\times \frac{1}{Z_{AB}} \int_{z_{AB-}}^{z_{AB+}} dz \frac{1}{z(1-z)} \int_0^{2\pi} \frac{d\phi}{2\pi} \Bigg[  (1-\Theta(G_{abk})) S_e^{\mbox{\tiny{QED}}}(\{p\},k; t)  \nonumber \\
&+ \Theta(G_{abk})\Bigg\{(1-\Theta((p^2_{\perp})_{ab})) S_e^{\mbox{\tiny{QED}}}(\{p\},k; t) \nonumber \\
&+ \Theta((p^2_{\perp})_{ab}) \bigg[\left(1 - \frac{a_e^{\mbox{\tiny{QED}}}((\{p\},k)_{ab})}{g(t)}\right) S_e^{\mbox{\tiny{QED}}}(\{p\},k; u) \nonumber \\
&+ \frac{a_e^{\mbox{\tiny{QED}}}((\{p\},k)_{ab})}{g(t)} \delta(\{p\},k - (\{p\},k)_{ab})\bigg] \Bigg\} \Bigg]
\end{align}
where
\begin{equation}
\Theta\left(G_{abk}\right) =
\left\{
	\begin{array}{ll}
		1  & \mbox{if } G_{abk} \geq 0 \mbox{ and } s_{ab} \geq 0\\
		0 & \mbox{else}
	\end{array}
\right.
\end{equation}
Note that the delta function in the last line of eq.~\eqref{emissionVeto} is written rather symbolically. From the algorithmic standpoint, it just means that the newly generated momenta are accepted. Using \eqref{emitPS} and taking the derivative with resect to $u$, we find the following differential equation
\begin{align}
\frac{\partial}{\partial u}S_e^{\mbox{\tiny{QED}}}(\{p\},k; u) &=  4\pi \alpha \sum_{\{ab\}} d\Phi^{\mbox{\tiny{ant}}}_{ab} \Theta((p^2_{\perp})_{ab}) \, \delta(u - t) \, a^e_{\mbox{\tiny QED}}((\{p\},k)_{ab}) \nonumber \\
&\times \bigg[\delta\left(\{p\},k - (\{p\},k)_{ab}\right) -  S_e^{\mbox{\tiny{QED}}}(\{p\},k; u)\bigg] 
\end{align}
which is solved by eq.~\eqref{QEDprob}. It is however not a unique solution. As is shown in \cite{Us}, the general class of solutions include a term that corresponds with a cutoff on $t$. Here, we leave it out for brevity.

Algorithm \ref{SVAemission} has multiple subsequent veto steps. From the above analysis, the probabilities of these veto steps are multiplicative. In the rest of this work, some additional veto steps will be introduced, of which one is the inclusion of a running coupling. For QED, we fix the value of the coupling at the electron mass and use the ordering variable of eq.~\eqref{QEDord} as renormalization scale. The scale-dependent coupling is then given by
\begin{equation} \label{running}
\alpha(t) = \frac{\alpha}{1 - \frac{\alpha}{3\pi}n_f(t) \log(\frac{t}{m_e^2})}
\end{equation}
where $n_f(t)$ is the number of active fermion flavors at scale $t$, weighted with the appropriate factors of charge and $N_c$. In the veto algorithm, running can be incorporated by using $\alpha(u)$ during generation of the shower variables and vetoing with probability 
\begin{equation}
P^{\mbox{\tiny{run}}} = \frac{\alpha(t)}{\alpha(u)}. 
\end{equation}
This factor could of course also be included in the emission kernel veto, but since that step is computationally the most expensive, some efficiency is gained by separating them and performing the cheaper veto steps first.

\subsection{Determining the overestimate} \label{overestimateSection}
Algorithm \ref{SVAemission} makes use of an overestimate function $g(t) \geq a_e^{\mbox{\tiny QED}}(\{p\},k)$. Due to the complex structure of the branching kernel of eq.~\eqref{aQED} and the definition of the ordering variable eq.~\eqref{tEmission}, the determination of this function is not easy. Since the singular structure of the branching kernel is regulated by the ordering variable, the overestimate should behave like 
\begin{equation} \label{overestForm}
g(t) = \frac{c}{t}
\end{equation}
where $c$ is a constant. Less singular terms can be added to affect behaviour for high values of $t$. A simple choice for $c$ which ensures that $g(t)$ overestimates the branching kernel is
\begin{equation} \label{simpleOverest}
c_{\mbox{\tiny{over}}} = 16 \sum_{\{ab\}} \max(0, -Q_aQ_b).
\end{equation}
This value can be found by discarding the contribution of the same-sign terms in eq.~\eqref{aQED} and realising that the largest opposite-sign contribution can be overestimated by $-16Q_aQ_b/t$. The branching kernel is exactly equal to this overestimate for events where all same-sign fermions are collinear with each other, and anticollinear to all opposite-sign fermions. However, eq.~\eqref{simpleOverest} overestimates the branching kernel in the majority of cases by a very large margin. This issue is particularly pronounced for high multiplicities, and significantly impacts the computation time. We therefore offer an alternative that should increase performance at the cost of introducing small fluctuations in weights for the events.

To improve on eq.~\eqref{simpleOverest}, properties of the pre-showering event have to be used. As it is difficult to find an upper limit of the branching kernel as a function of the pre-branching event, we resort to using a value for $c$ such that eq.~\eqref{overestForm} is an incomplete overestimate. This means that it overestimates the branching kernel in the majority of phase space, but falls short in some small regions. The resulting discrepancy is corrected by introducing small deviations in the weights of the events.  

We make use of a modified version of the Sudakov veto algorithm that has previously been used in \cite{Reweighting1,Reweighting2,Reweighting3} to estimate shower uncertainties. It is given in algorithm \ref{SVAreweighting} for a general single-variable case to avoid notational cluttering. The function $g(t)$ is an incomplete overestimate of the branching kernel $f(t)$, which means that their ratio $r(t) = f(t)/g(t)$ cannot used as the veto probability. Instead, a veto probability $p(r(t))$ is introduced and corrected for in the weights. Algorithm \ref{SVAreweighting} can easily be shown to produce the desired distribution using the methods shown in the previous section and \cite{Us}. 
\begin{algorithm} 
\caption{The Sudakov veto algorithm incomplete overestimates}
\label{SVAreweighting}
\begin{algorithmic}
\STATE $t \leftarrow u$
\STATE $w \leftarrow 1$
\LOOP
\STATE $t \leftarrow$ solution of $\rho_1 = \exp\left(-\int_{t'}^t d\tau \, g(\tau) \right)$ for $t'$
\IF {$\rho_2 < p(r(t))$}
\STATE $w \leftarrow w \frac{r(t)}{p(r(t))}$
\RETURN $t, w$
\ELSE
\STATE $w \leftarrow \frac{1-r(t)}{1-p(r(t))} w$
\ENDIF
\ENDLOOP
\end{algorithmic}
\end{algorithm}
A sensible choice for $p(r(t))$ should provide better performance while maintaining small fluctuations in weights. This can be achieved by choosing a function which behaves as closely as possible to $r(t)$ for ratios between zero and one, where the overestimate $g(t)$ truely overestimates $f(t)$, while moving close to one as $r(t) > 1$. A possible choice is 
\begin{equation} \label{sigmoid}
p(r(t)) = \tanh(r(t)).
\end{equation}

\begin{figure}[!htb]
\centering
\includegraphics[scale=0.8]{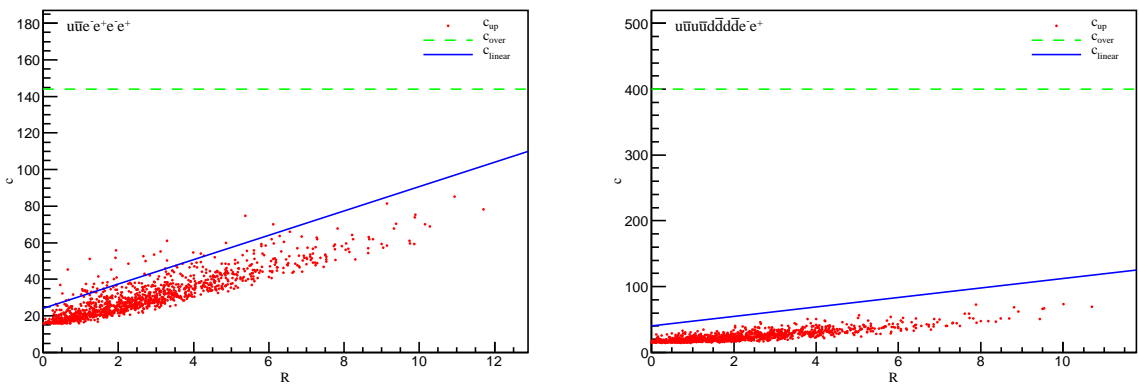}
\caption{The correlation between $R$ and the upper limit of the branching kernel. To approximately determine these upper limits, shower emissions are sampled from events generated uniformly with the RAMBO algorithm \cite{Rambo}. The value of $c_{\mbox{\tiny{up}}}$ is the required value of the overestimate constant $c$ found by evaluating the branching kernel for every sampled emission. The dashed green line corresponds with the absolute overestimate given by eq.~\eqref{simpleOverest}. The solid blue line represents the incomplete overestimate given by eq.~\eqref{cLinear}.}
\label{Rcorrelation}
\end{figure}

Note that the weights of the events should remain very close to unity with this choice, but there is a tiny probability to produce negatively weighted events when a scale is generated in a region where $r(t)>1$, but gets rejected. Next, we search for a relation between the pre-branching event and the upper limit of the branching kernel. One variable that is correlated with this upper limit is
\begin{equation}
R = - \sum_{\{ab\}} Q_{a} Q_{b} (1 - \cos(\theta_{ab})),
\end{equation}

where $\theta_{ab}$ is the angle between the momenta $p_a$ and $p_b$. The variable $R$ resembles the eikonal factors without the photon energy, if they were evaluated with pre-branching fermion momenta. The correlation with the upper limit of the branching kernel is shown in figure \ref{Rcorrelation} for some final states. The values $c_{\mbox{\tiny{up}}}$ are the approximate upper limits of the branching kernels of individual events. The incomplete overestimate 
\begin{equation} \label{cLinear}
c_{\mbox{\tiny{linear}}} = 4n + 8\left(1 - \frac{4n}{c_{\mbox{\tiny{over}}}}\right)R,
\end{equation}
where $n$ is the number of charged fermions, is chosen to coincide with $c_{\mbox{\tiny{over}}}$ at $r_{\mbox{\tiny{max}}} = c_{\mbox{\tiny{over}}}/8$. In section \ref{performanceTesting}, the performance of algorithm \ref{SVAreweighting} with eq.~\eqref{cLinear} is tested.

\subsection{Ordering} \label{orderAndMerge}
An ordering strategy is a necessary ingedient for a parton shower to produce properly resummed event samples. It serves as a means to restrict the available Monte Carlo pathways to a phase space point, such that overcounting is prevented and the leading contributions are kept. For QCD antenna showers it has been shown in \cite{vinciaMassless,vinciaMassive} that the absence of ordering produces a substantial overcounting in the hard, wide-angle region. We perform a comparable analysis for QED radiation in section \ref{MECsection}. The VINCIA parton shower implements two\footnote{Recently, the VINCIA shower discontinued the use of smooth ordering and now only employs strong ordering.} ordering strategies that are closely connected to the method of merging the shower with exact matrix element calculations. These ordering strategies can be shown to produce the same leading-log behaviour \cite{vinciaHadron}. 

Strong ordering is the traditional way of ordering branchings in a parton shower. An additional veto
\begin{equation}
\Theta(t_{\mbox{\tiny{prev}}} - t)
\end{equation}
is applied in the Sudakov veto algorithm, where $t_{\mbox{\tiny{prev}}}$ is the scale of the previous shower branching. The application of this step function is equivalent to restarting the shower from $t_{\mbox{\tiny{prev}}}$ for subsequent branchings. The inclusion of this veto adds an explicitly non-Markovian element to the parton shower. In addition, unpopulated zones appear in the multi-branching phase space when there is no ordered path to reach them. In section \ref{MECsection} these zones are shown for several final state configurations. These properties of strong ordering have consequences for the implementation of the unitary matching and merging method employed by VINCIA \cite{orderedMerging}. 

An alternative is given by smooth ordering. Subsequent emissions are instead allowed to cover the entire available phase space. An ordering criterion is introduced by means of an additional veto in the Sudakov veto algorithm with probability 
\begin{equation} \label{Pord}
P^{\mbox{\tiny{ord}}} = \frac{t}{t + \hat{t}},
\end{equation}
where $\hat{t}$ is the 'current' scale of the event. It is defined as the minimum of the scales that correspond with all available shower clustering of the pre-branching event. This determination of the scale accounts for all possible ways at which the shower could have reached the pre-branching state, and is therefore entirely Markovian. Smooth ordering offers the distinct advantage of covering the entire phase space with every emission. Unitary matching and merging is therefore more straightforward \cite{vinciaMassless}. However, it was noted in \cite{newVinciaVersion} that the Sudakov factors in the unordered regions are probably not correct. 

\section{Photon splitting} \label{splittingSection}
From the parton shower perspective, photon splitting is much simpler than photon emission. No soft singularity is present in the photon splitting kernels, so their treatment is very similar to the QCD counterpart. As we will explain, the only significant difference is the absence of a color structure to aid in the selection of a secondary participating particle for the photon splitting. We again first define the splitting kernel and proceed with the shower implementation.    

\subsection{Splitting kernel}
Let $|\bar{M}_1(\{p\}, p_a, p_b)|^2$ be the squared amplitude of some process, where $p_a$ and $p_b$ are the momenta of a charged fermion-antifermion pair. This matrix element factorizes in the quasi-collinear limit for $p_a$ and $p_b$ as defined by eq.~\eqref{quasiCollinear}.
\begin{equation} \label{splitFac}
|M_1(\{p\}, p_a, p_b)|^2 =  4\pi\alpha \, Q_f^2 \frac{2}{m_{ab}^2} P(z, m_f^2, m_{ab}^2) |M_0(\{p\},p_k)|^2
\end{equation}
where $m_f$ and $Q_f$ are the mass and charge of the fermion-antifermion pair and
\begin{equation}
P(z, m_f^2, m_{ab}^2) = z^2 + (1-z)^2 + 2\frac{m_f^2}{m_{ab}^2}.
\end{equation}
Since there are no additional soft properties, the factorization is unrelated to a third particle. However, an additional \emph{spectator} particle is still required for the kinematics. We therefore derive the splitting kernels from matrix elements including these spectators. For a photon spectator, we make use of an effective $H \gamma \gamma$ coupling, while for a fermionic spectator we make use of an effective $\mu \bar{e} \gamma$ coupling. As is to be expected, the splitting kernels turn out to be the same up to nonsingular terms. They are
\begin{align} \label{QEDsplitAntenna}
a_{s}^{\mbox{\tiny{QED}}}(p_a, p_b, p_c) &= \frac{|M(H \rightarrow f \bar{f} \gamma)|^2} {4\pi \alpha |M(H \rightarrow \gamma \gamma)|^2} + \mathcal{O}(1) = \nonumber \frac{|M(\mu \rightarrow e f \bar{f})|^2}{4\pi \alpha |M(\mu \rightarrow e \gamma)|^2} + \mathcal{O}(1) \nonumber \\
&= \, Q^2_f\frac{2}{m_{ab}^2} \bigg[ \frac{s_{ac}^2 + s_{bc}^2}{S_{KC}^2} + 2\frac{m_f^2}{m_{ab}^2}\bigg]
\end{align}
where $p_c$ is the spectator momentum.

\subsection{Ordering variable and phase space} \label{splitAriSection}
\begin{figure}
\centering
\includegraphics[width=8cm]{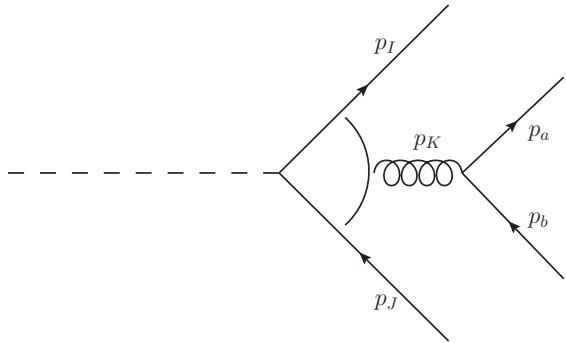}
\caption{A shower history where a gluon splitting follows a gluon emission.}
\label{ariFigure}
\end{figure}

We denote the participating momenta as
\begin{equation}
p_K + p_C \rightarrow p_a + p_b + p_c
\end{equation}
where $p_a^2 = p_b^2 = m_f^2$, $p_C^2 = p_c^2 = m_c^2$ and $p_K^2=0$. Since the only singularity is of collinear nature, it is sufficient to use the invariant mass of the produced fermion-antifermion pair as the ordering variable. We follow VINCIA and use the shower variables
\begin{equation}
t = m^2_{ab} \qquad z = \frac{s_{bc}}{m^2_{KC}}.
\end{equation}
The massless boundaries on $z$ now are
\begin{equation}
0 \leq z \leq 1 - \frac{t}{m_{KC}^2}.
\end{equation}
Transforming the antenna phase space to these shower variables leads to
\begin{equation} \label{splitPS}
d\Phi_{KC}^{\mbox{\tiny{ant}}} = \frac{1}{16 \pi^2} dt \, dz \frac{d\phi}{2\pi} \frac{m^2_{KC}}{m_{KC}^2 - m_c^2} \theta(m_{CK}^2-s_{ab}-s_{bc} - m_c^2 - 2m_f^2 ) \theta(G_{abc}>0).
\end{equation}
To determine the shower approximation, a method of selecting the spectator particle is required. At first glace, this choice does not seem very significant and one might be tempted to select a spectator at random. However, as will be shown in section \ref{MECsection}, this can lead to a significant overcounting of the matrix element. Instead, we will generalize a method used by ARIADNE and VINCIA for gluon splitting. In leading-color QCD, the available spectators are the two partons which are color-adjacent to the splitting gluon. Labelling them $I$ and $J$ and the gluon $K$, the probability to select parton $I$ as spectator is given by 
\begin{equation} \label{ariFacQCD}
P_{KI}^{\mbox{\tiny{Ari}}} = \frac{m_{KJ}^2}{m_{KI}^2 + m_{KJ}^2}.
\end{equation}
 To see why, consider the shower history sketched in figure \ref{ariFigure}. Both the emission and the splitting are performed by the parton shower. This means that the gluon is on-shell after the emission. If the gluon is collinear with one of the fermions, say $p_I$, then the invariant mass $m_{IK}^2$ is small. If $p_I$ is selected as the spectator for the splitting of $p_K$ into $p_a$ and $p_b$, the invariant mass $m_{IK}^2$ remains unchanged, but if $p_K$ is selected, $m_{IK}^2$ can become large. Therefore, the small value that was used in the emission kernel was incorrect by a significant margin. The selection probability of eq.~\eqref{ariFacQCD} gives preference to spectators which have low invariant mass with $p_K$, suppressing this effect.

For QED, eq.~\eqref{ariFacQCD} needs to be expanded to be able to account for more than two candidate spectators. A suitable choice is 
\begin{equation} \label{ariFacQED}
P_{KI}^{\mbox{\tiny{Ari}}} = \frac{1/m_{IK}^2}{\sum_{J} 1/m^2_{JK}}
\end{equation} 
where $J$ now runs over all available spectators. It is easy to see that this definition satisfies the requirements explained above, and that it reduces to eq.~\eqref{ariFacQCD} for two spectator candidates. The shower approximation is
\begin{equation}\label{QEDsplitApprox}
|M_{n+1}(\{p\},p_a,p_b)|^2 \approx \sum P_{KC}^{\mbox{\tiny{Ari}}} a^{s}_{\mbox{\tiny{QED}}}(p_a, p_b, p_c) |M_{n}(\{p\},p_K)|^2
\end{equation}
and the corresponding targeted shower probability distribution should be
\begin{align} \label{splitProb}
S_s^{\mbox{\tiny{QED}}}(\{p\},p_a,p_b) &= \sum_{K} \sum_{C} P_{KC}^{\mbox{\tiny{Ari}}} a_{s}^{\mbox{\tiny{QED}}}(p_a, p_b, p_c)  \theta(u - t) \nonumber \\
&\times \exp\left( \sum_{K'}\sum_{C'} \int_t^u d\Phi_{K'C'}^{\mbox{\tiny{ant}}} P_{K'C'}^{\mbox{\tiny{Ari}}} a_{s}^{\mbox{\tiny{QED}}}(p'_a, p'_b, p'_c) \right).
\end{align}

\subsection{Kinematics}
We employ a simplified kinematic strategy for photon splitting as compared with emissions, since the map only needs to account for collinear safety. The pre-branching and post-branching momenta are related by
\begin{align} \label{splittingMap}
p_K &= x(p_a + p_b) + zp_c  \nonumber \\
p_C &= (1-x)(p_a + p_b) + (1-z)p_c.
\end{align}
The momenta $p_a$ and $p_b$ are multiplied with the same parameter such that the collinear limit corresponds with $x \rightarrow 1$ and $z \rightarrow 0$. Solving $p_k^2 = 0$ and $p_A^2 = m_a^2$ fixes $x$ and $z$. They are given by
\begin{align}
x &= \frac{m_{KC}^2 - m_c^2}{2m_{KC}^2} \left(1 + \frac{2m_c^2 + (s_{ac} + s_{bc})}{\sqrt{r}} \right) \nonumber \\
z &= \frac{m_{KC}^2 - m_c^2}{2m_{KC}^2} \left(1 - \frac{2m_{ab}^2 + (s_{ac} + s_{bc})}{\sqrt{r}} \right)
\end{align}
where 
\begin{equation}
r = (s_{ac} + s_{bc})^2 - 4m_c^2 m_{ab}^2
\end{equation}
Note that in the massless limit $x\rightarrow 1$ and $p_C$ only recoils longitudinally. Then, in the collinear limit $s_{ab} \rightarrow 0$, $z \rightarrow 0$ making the mapping collinear safe. Computing the angle between $p_C$ and $p_c$, we find
\begin{equation}
\cos\psi = \frac{E_A E_a - (1-z)m_a^2 - (1-x)(s_{ab} + s_{ac})}{|\vec{p}_A||\vec{p}_a|}
\end{equation}
which fixes the kinematics entirely.

\subsection{Sudakov veto algorithm}
Since for photon splitting, the splitting kernel can be partitioned into independently radiating antenna functions corresponding with the choice of the photon and the specator, the Sudakov veto algorithm for photon splitting is much closer to the standard algorithms used in QCD. We again first define the overestimates for the boundaries of $z$
\begin{equation}
z_{KC-} = 0 \quad z_{KC+} = 1 - \frac{t_{\mbox{\tiny{cut}}}}{m^2_{KC}}
\end{equation}
and channel-specific weights
\begin{equation}
w_{KC} = \frac{1}{4\pi} z_{KC+} \frac{m_{KC}^2}{m_{KC}^2 - m_c^2} P_{KC}^{\mbox{\tiny{Ari}}} \qquad w = \sum_{K} \sum_{C} W_{KC}.
\end{equation}
The required overestimate for the antenna function is easily derived. Since eq.~\eqref{QEDsplitAntenna} is only singular in the invariant of the produced fermion-antifermion pair and the terms inside the brackets are easily overestimated with constants, the most sensible choice is
\begin{equation}\label{splittingOverest}
f(t) = \frac{4}{t}
\end{equation}
where we made use of $Q_f^2 < 1$ for all Standard Model fermions. Note that this overestimate is independent of the selected photon or spectator. The Sudakov veto algorithm for photon splitting is given by algorithm \ref{SVAsplitting}. As written, the algorithm only allows for a single flavor of the fermion-antifermion pair. Due to the independence of the overestimate on the selected channel, including additional flavors is straightforward. For $n_f$ flavors, the weights should be adjusted according to 
\begin{equation}
w_{KC} \rightarrow n_f w_{KC}
\end{equation}
and the algorithm should be modified to select a flavor at random. Even for massive flavors, the probabilities are adjusted accordingly by the veto step. 
\begin{algorithm} [H]
\caption{The Sudakov veto algorithm for photon splitting}
\label{SVAsplitting}
\begin{algorithmic}
\STATE $t \leftarrow u$
\LOOP
\STATE Choose a photon $K$ and a spectator $C$ with probability $w_{KC}/W$
\STATE $t \leftarrow$  solution of $\rho_1 = \exp\left(-\int_{t'}^t d\tau \, W \alpha f(\tau) \right)$ for $t'$
\STATE $z \leftarrow z_+ \rho_2$ 
\STATE $\phi \leftarrow 2\pi \rho_3$
\STATE Compute $s_{ab}$, $s_{ac}$, and $s_{bc}$ from $t$, $z$, and $m_{KC}^2$
\IF {$G_{abc} > 0$ and $s_{ac} > 0$}
\STATE Construct the $p_a$, $p_b$ and $p_c$ from $p_C$, $p_K$, $t$, $z$ and $\phi$
\IF {$\rho_4 < a_s^{\mbox{\tiny QED}}(p_a, p_b, p_c) / f(t)$}
\RETURN $p_a$, $p_b$ and $p_c$
\ENDIF
\ENDIF
\ENDLOOP
\end{algorithmic}
\end{algorithm}
We again analyze algorithm \ref{SVAsplitting} to show that it produces the density given by eq.~\eqref{splitProb} by explicitly writing out the probabilities.
\begin{align}
S_s^{\mbox{\tiny{QED}}}(\{p\},p_a,p_b; u) &= \frac{1}{W} \sum_{K} \sum_{C} w_{KC} \int^u_0 dt W \alpha f(t) \exp\left(-\int^u_t d\tau W \alpha f(\tau) \right) \nonumber \\
&\times \frac{1}{z_{KC+}} \int_0^{z_{KC+}} dz \int_0^{2\pi} \frac{d\phi}{2\pi} \Bigg[(1 - \Theta(G_{abc})) S^s_{\mbox{\tiny{QED}}}(\{p\},p_a,p_b; u) \nonumber \\
&+ \Theta(G_{abc}) \Bigg\{\left(1 - \frac{a_s^{\mbox{\tiny QED}}(p'_a, p'_b, p'_c)}{f(t)} \right)S_s^{\mbox{\tiny{QED}}}(\{p\},p_a,p_b; t)  \nonumber \\
&+ \frac{a_s^{\mbox{\tiny QED}}(p'_a, p'_b, p'_c)}{f(t)} \delta(\{p\},p_a,p_b - p'_a, p'_b,p'_c)\Bigg\}\Bigg]
\end{align}
where again the delta function is written symbolically. The newly generated momenta are primed to distinguish them from the arguments of the probability distribution. Taking the $u$-derivative leads to
\begin{align}
\frac{\partial}{\partial u}S_s^{\mbox{\tiny{QED}}}(\{p\},p_a,p_b; u) &= 4\pi\alpha\sum_{K} \sum_{C} d\Phi_{KC}^{\mbox{\tiny{ant}}} P_{KC}^{\mbox{\tiny{Ari}}} \delta(u-t) \, a_s^{\mbox{\tiny QED}}(p'_a, p'_b, p'_c) \nonumber \\
&\times \bigg[\delta(\{p\},p_a,p_b - p'_a, p'_b,p'_c) - S_s^{\mbox{\tiny{QED}}}(\{p\},p_a,p_b; u) \bigg]
\end{align}
which is solved by eq.~\eqref{splitProb}.

\section{Validation} \label{resultsSection}
In this section, we compare the QED shower method described in the previous sections to theory results. We first compare the shower approximation to fixed order matrix elements to inverstigate how it performs outside the singular regions. Next, we verify the validity of the sector strategy for photon emissions by comparing the shower implementation to a numerical solution of the DGLAP equation. Then, we compare to the YFS formalism used in \cite{decays1,decays2} to validate the soft behaviour of the shower. The potential issue of discontinuities in the emission phase space is discussed, and the performance of the method described in section \ref{overestimateSection} is tested. 

\subsection{Comparison with matrix elements} \label{MECsection}
In this section, we test the shower approximations given by eq.~\eqref{QEDshowerApprox} and eq.~\eqref{QEDsplitApprox}. The shower approximations are only exact in their corresponding singular regions, where most radiation is produced. However, the shower populates a much larger part of the available phase space where equations~\eqref{QEDshowerApprox} and~\eqref{QEDsplitApprox} are only approximately valid. To gain some insight in the quality of the shower approximations we compare them to fixed-order matrix elements, varying the types and number of branchings, the type of ordering and the inclusion of the selection probability given by eq.~\eqref{ariFacQED}.  

Similar to VINCIA \cite{vinciaMassless,vinciaMassive}, we compare to fixed order calculations by selecting some process with an $n$-particle final state and sampling the $n$-body phase space uniformly using the RAMBO algorithm \cite{Rambo}. The fixed-order matrix element $|M_n|^2$ is computed using MadGraph5 \cite{Madgraph5}. The parton shower approximation is applied multiple times until an $m$-particle final state matching matrix element $|M_m|^2$ is reached. To achieve this, the sampled momenta are clustered through eq.~\eqref{emissionMap} and eq.~\eqref{splittingMap}, inverting the normal parton shower process. In most cases, the parton shower can reach a phase space point through multiple paths which all contribute to the Monte Carlo probability. In this comparison, all possible shower histories are therefore summed over. The ratio between the parton shower and the fixed order calculation is then computed as
\begin{equation} \label{MEratio}
\frac{PS}{ME} = \frac{\sum_{\mbox{\tiny{histories}}}a_n^{\mbox{\tiny{QED}}} P_n ... a_{n-m}^{\mbox{\tiny{QED}}}P_{n-m}|M_m|^2}{|M_n|^2}
\end{equation}
where $a_i^{\mbox{\tiny{QED}}}$ is the branching kernel for the $i$-th clustering, which can be both emissions and splittings. The term $P_n$ contains additional factors including eq.~\eqref{ariFacQED} for photon splittings and an ordering factor, which is a step function in case of strong ordering or eq.~\eqref{Pord} in case of smooth ordering. We keep $\alpha$ constant everywhere such that it drops out in eq.~\eqref{MEratio}.

\subsubsection{Photon emission}
The sector approach to photon emissions is only different from the normal dipole or antenna shower strategy if more than two charged fermions are involved. We therefore consider decays to final states of $4$ and $6$ charged fermions. In the Standard Model, decays like $H \rightarrow 4l$ correspond with highly structured matrix elements which should not be probed in this comparison. Instead, we add a scalar $\phi$ to the Standard Model which directly couples to either $4$ or $6$ charged leptons. This causes the comparison to mostly probe the parton shower approximation only. However, we stress that the following results do not offer a direct representation of the accuracy of the shower. Not only do the results depend on the underlying process, but the parton shower does not sample phase space uniformly as is done in this comparison, but rather it prefers the singular regions where it performs best. Instead, these comparisons offer insight into the impact of algorithmic choices such as the type of ordering or the spectator selection probability given by eq.~\eqref{ariFacQED}. 

\begin{figure}[!htb]
\centering
\hspace*{-0.8cm}\includegraphics[scale=0.8]{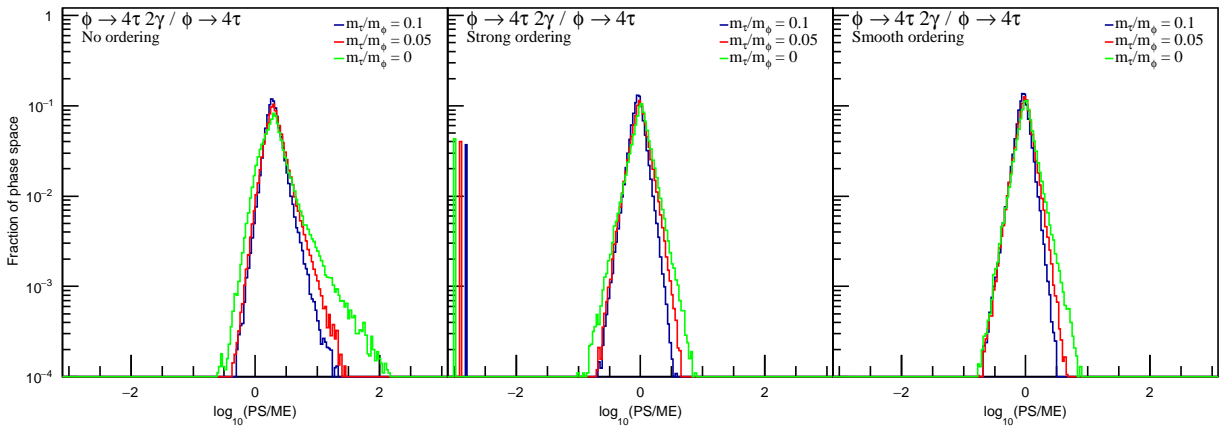}
\caption{Comparison of the shower approximation to matrix elements for $\phi \rightarrow 4\tau 2\gamma$ matched to $\phi \rightarrow 4\tau$ for different types of ordering. The bars on the left side of the strong ordering comparison correspond with events that lie in the dead zone of the shower.}
\label{emissionOrderingME}
\end{figure}

We first plot the ratio given by eq.~\eqref{MEratio} in figure \ref{emissionOrderingME} for a scalar $\phi$ decaying to four leptons and two photons. The matching matrix element is the decay of $\phi$ to four leptons, so the showering component only consists of two emissions. Figure \ref{emissionOrderingME} illustrates the impact of the types of ordering discussed in section \ref{orderAndMerge}.
From the left-hand plot, it is clear that the matrix element is significantly overestimated by the shower approximation. For the process at hand, the only two available shower histories are defined by the order in which the photons are clustered. When no ordering condition is imposed, both paths contribute to the shower approximation, overestimating the matrix element. When strong ordering is imposed, one of these paths will in most cases not contribute. However, occasionally either both paths or no paths will contribute. This is caused by the changing fermion momenta after the primary clustering, changing the ordering scale of the secondary clustering. In the middle graph of figure \ref{emissionOrderingME}, all phase space points where neither path contributes are contained in the bars on the left side. These events constitute the dead zone in the parton shower phase space. For smooth ordering, the paths are instead weighted by a continuous ordering probability, preventing the occurence of a dead zone. 
\begin{figure}[!htb]
\centering
\hspace*{-0.8cm}\includegraphics[scale=0.8]{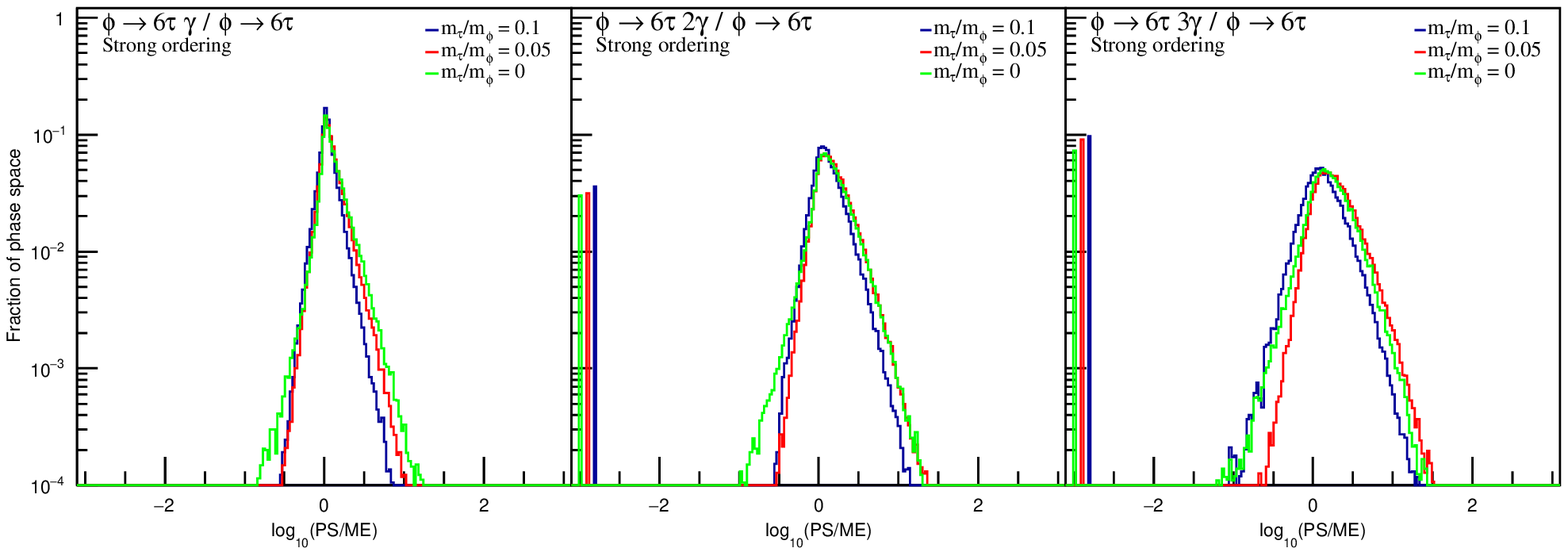}
\hspace*{-0.8cm}\includegraphics[scale=0.8]{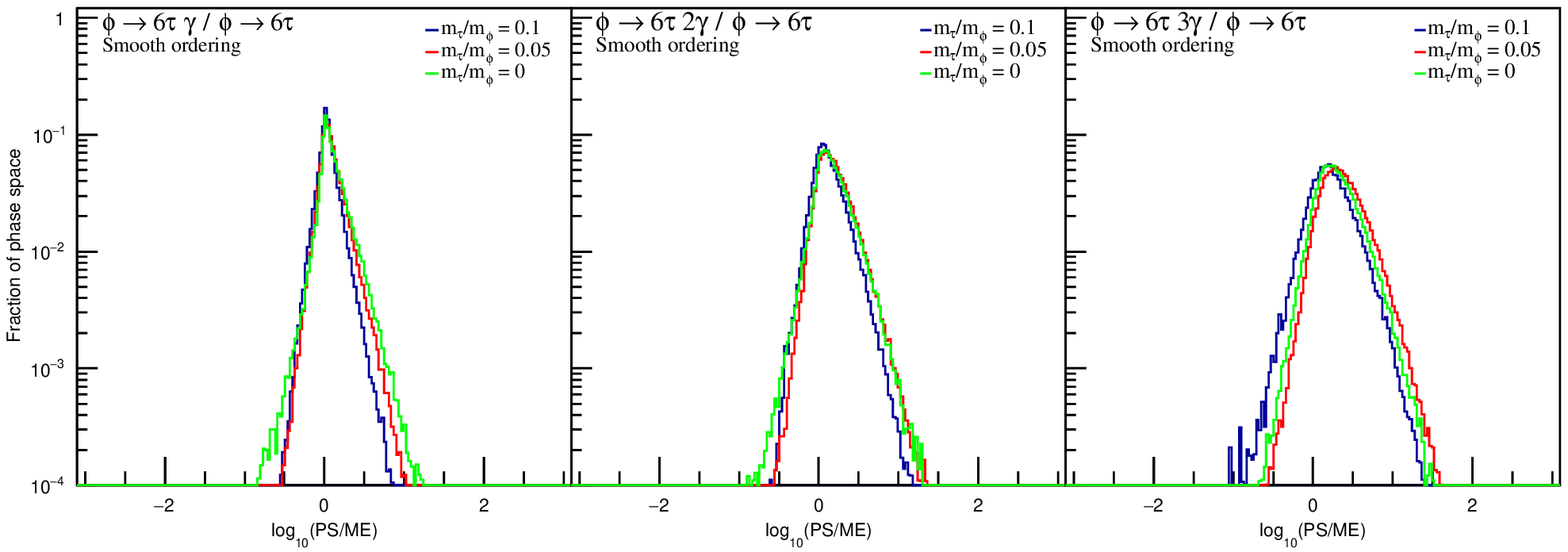}
\caption{Comparison of the shower approximation to matrix elements for one, two and three emissions from $\phi \rightarrow 6\tau$ using strong and smooth ordering.}
\label{emissionME}
\end{figure}

In figure \ref{emissionME} the comparison to matrix elements is shown for one, two and three emissions from $\phi \rightarrow 6\tau$. The shower approximation appears to perform better for more massive leptons, but this is caused by the decreased phase space available for the photon, pushing it to be softer more often. Finally, the graphs are somewhat shifted in the higher multiplicity cases. This is a direct consequence of the absence of mass-dependent non-singular terms in the emission kernel, which become more relevant for higher lepton masses.

\subsubsection{Photon splitting}
\begin{figure}[!htb]
\centering
\hspace*{-0.8cm}\includegraphics[scale=0.8]{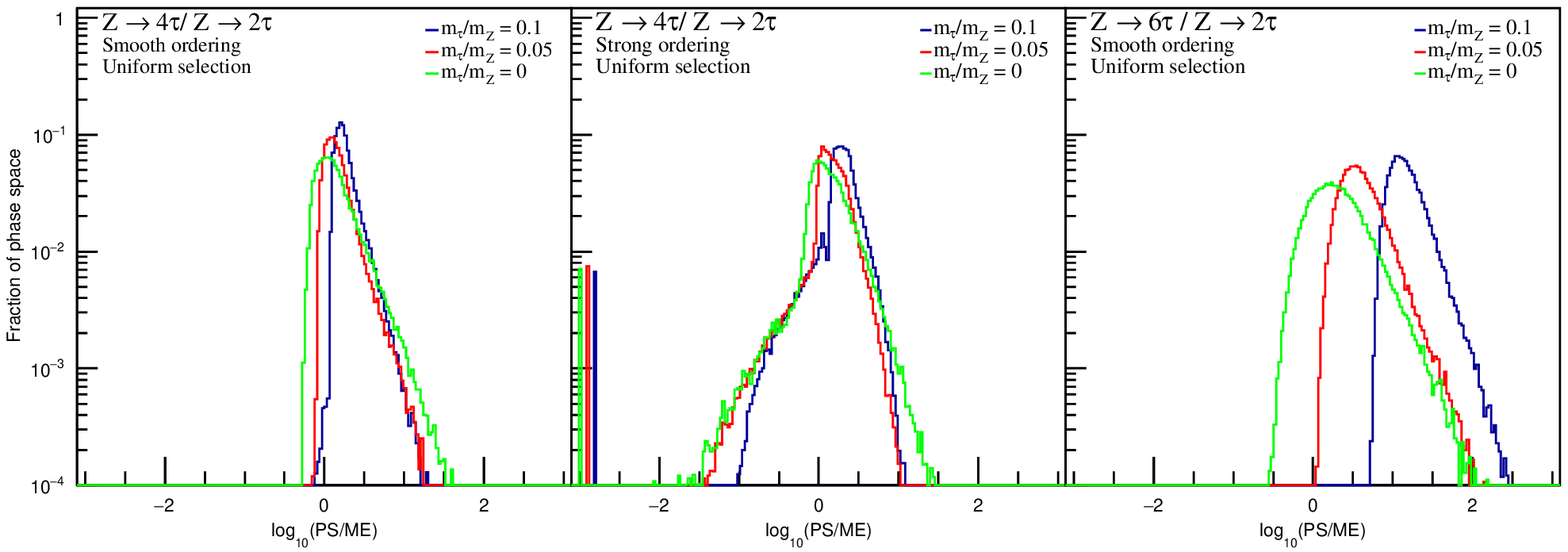}
\hspace*{-0.8cm}\includegraphics[scale=0.8]{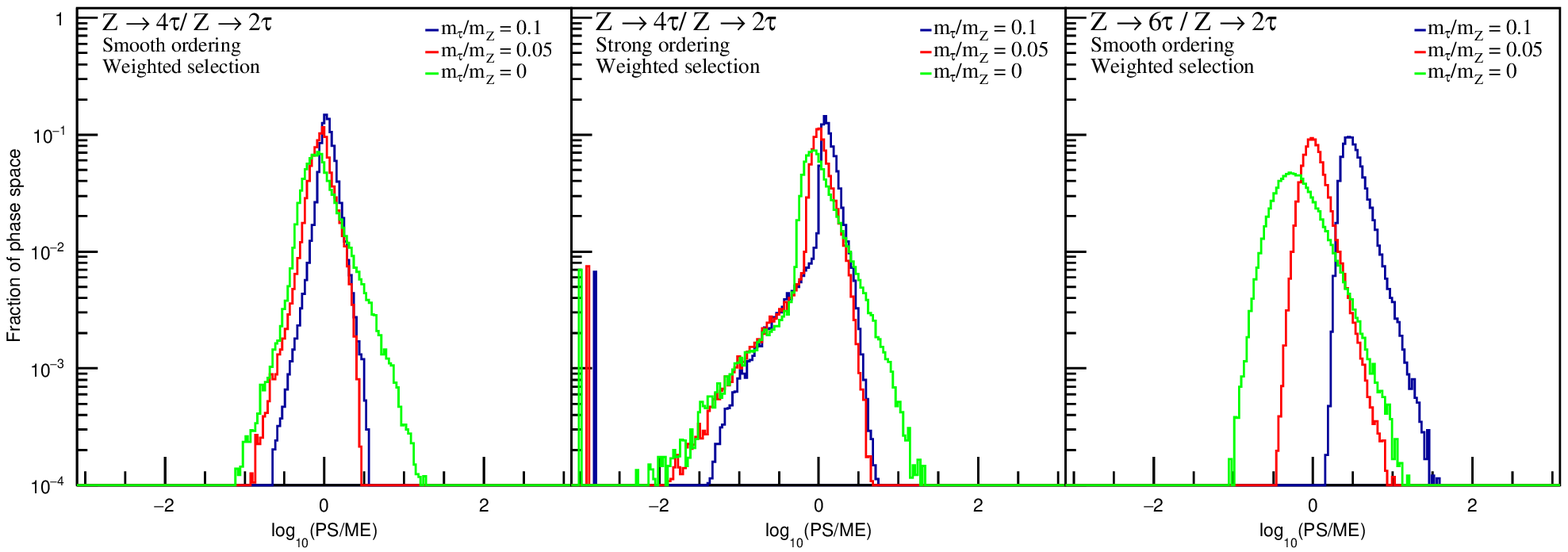}
\caption{Comparison of the shower approximation to matrix elements for a combination of emissions and splittings from $Z \rightarrow 2\tau$. In the top row, the spectator is selected uniformly, while in the bottom row, the selection probability given by eq.~\eqref{ariFacQED} is used.}
\label{ariME}
\end{figure}

We now incorporate photon splitting to massless and massive lepton-antilepton pairs. As discussed in section \ref{splitAriSection}, the parton shower approximation can be improved by using a weighted selection of a spectator for a photon splitting, instead of choosing uniformly. To check this, we compare processes involving emissions and splittings from the decay $Z \rightarrow 2\tau$. In figure \ref{ariME} the shower approximation is compared to matrix elements for a combination of emissions and splittings. In the top row, the spectator is selected uniformly, while in the bottom row it is selected with probability given by eq.~\eqref{ariFacQED}. The weighted selection strategy improves the smoothly ordered results more than the strongly ordered result. We have checked this to be true for multiple matrix elements. In \cite{vinciaMassive} it was shown that the agreement between the matrix element and the parton shower approximation depends on the combination of the choice of ordering variable and the method of selecting a spectator. Until the QED shower is interleaved with a QCD shower, the effect of the choice of ordering variable remains ambiguous and we prefer to maintain the choice made in VINCIA due to the similarity between gluon and photon splitting. 

We also note that more significant shifting occurs for the higher multiplicity processes as compared with the case of pure emissions. The photon splitting kernel is singular only in the (quasi-)collinear limit, and even there only single poles occur. Photon emissions are instead associated with double poles. It is therefore to be expected that the influence of the process-specific non-singular terms increases significantly for photon splittings, worsening the quality of the parton shower approximation.

\begin{figure}[!htb]
\centering
\hspace*{-0.8cm}\includegraphics[scale=0.8]{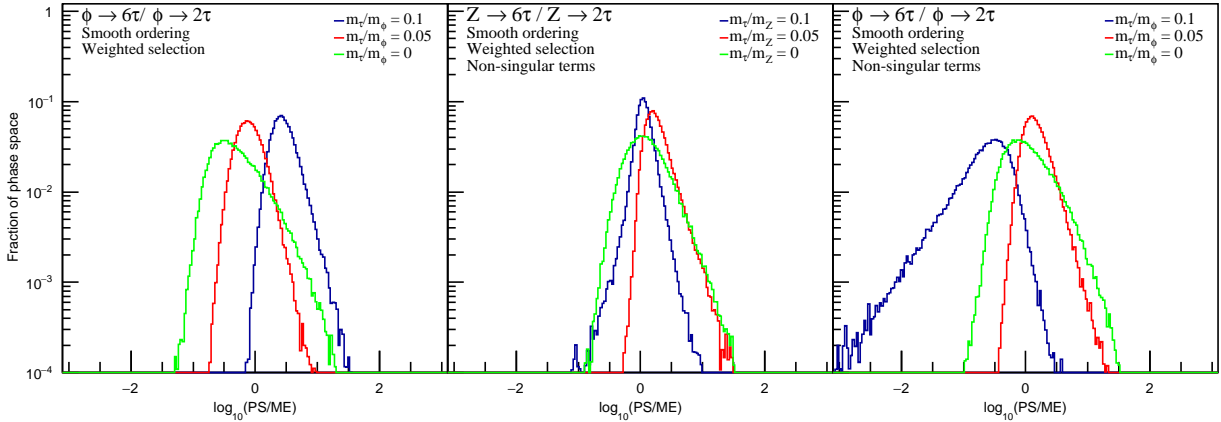}
\caption{Illustration of the impact of non-singular terms on the comparison between the parton shower approximation and matrix element calculations for two emissions and two splittings. In the left-hand graph, the $Z$-boson is replaced with a scalar that decays to a fermion-antifermion pair. In the middle and right-hand graph, the same comparison is repeated for the decay from both a $Z$ and a $\phi$, but the splitting antenna function now includes non-singular terms.}
\label{nonSing}
\end{figure}

In figure \ref{nonSing}, the impact of non-singular terms is illustrated. In the middle and right-hand plot, the splitting kernel is modified to 
\begin{equation} \label{splitWithNonSing}
a_{s}^{'\mbox{\tiny{QED}}}(p_a, p_b, p_c) = Q^2_f\frac{2}{m_{ab}^2} \bigg[ \frac{s_{ac}^2 + s_{bc}^2}{S_{KC}^2} + 2\frac{m_f^2}{m_{ab}^2}\bigg] + Q_f^2 \bigg[10 \frac{1}{m_{Z}^2} - 1500 \frac{m_f^2}{m_Z^4}\bigg]
\end{equation}
where $m_Z$ only serves as a means to fix the dimensionality of the non-singular terms. The coefficients are loosely chosen to show that the peaks in the middle graph of figure \ref{nonSing} can be aligned and centralized. The same non-singular terms are used in the right-hand pane, showing that they are not a universal improvement. 

\subsection{Comparison with analytic resummation}
For many exclusive observables, large logarithmic corrections of the form $\alpha^n \ln^n\left(Q^2/m^2\right)$ or $\alpha^n \ln^n\left(Q^2/t_{\mbox{\tiny{cut}}} \right)$ appear in cross section calculations at perturbative order $n$. These logarithms are a direct consequence of the singular regions the parton shower aims to cover correctly. Here, $Q^2$ is the hard squared scale of the process, and the singular behaviour leading to these logarithms is regulated by either $t_{\mbox{\tiny{cut}}}$ for massless particles, or $m^2$ for particles of mass $m$. Since these logarithmic contributions can be sizeable at every order in perturbation theory, they have to be resummed to all orders to obtain reliable results. Resummation can be achieved analytically or numerically using a parton shower. In this section, we compare the sector approach to photon emissions with results from analytical resummation, validating that it produces the correct collinear logarthms. For simplicity, we restrict the comparison to massless leptons.

In QCD, the evolution from high to low scales of the final state parton energy is given by the partonic fragmentation functions. While in QCD these functions are related to hadronization, a similar concept can be introduced for leptons. Naming these functions $L(x,t)$, they describe the distribution for a lepton to retain a fraction $x$ of its original energy at an energy squared scale $t$ which is lower than the hard scale $Q^2$. Note that these distributions are not sensitive to soft wide-angle radiation and should therefore also be reproduced by incoherent showers. The function $L(x,t)$ is completely analogous to the QCD fragmentation function, and thus also obeys the DGLAP equation
\begin{equation} \label{DGLAP}
t \frac{\partial}{\partial t} L\left(x,t\right) = \int_x^1 \frac{\alpha(t)}{2\pi} \frac{dz}{z} P_{ll}(z) L\left(x/z,t\right)
\end{equation}
where $P_{ll}(z)$ is the regularized Altarelli-Parisi splitting function
\begin{equation}
P_{ll}(z) = \frac{2}{(1-z)_+} + \frac{3}{2}\delta(1-z) - (1+z).
\end{equation}
Usually, a transformation to Mellin space is used to turn eq.~\eqref{DGLAP} into an ordinary linear differential equation. We instead opt to solve it numerically using the methods described in \cite{numericalDGLAP}. 

To compare with the shower approach, we start from a RAMBO-generated four-lepton system. The definition of $x$ is
\begin{equation}
x = \frac{E_{t_{\mbox{\tiny{cut}}}}}{E_{Q^2}}
\end{equation}
where $E_{Q^2}$ and $E_{t_{\mbox{\tiny{cut}}}}$ are the energies of one of the leptons at respectively the hard scale and the cutoff scale. 

\begin{figure}[!htb]
\centering
\hspace*{-0.8cm}\includegraphics[scale=0.8]{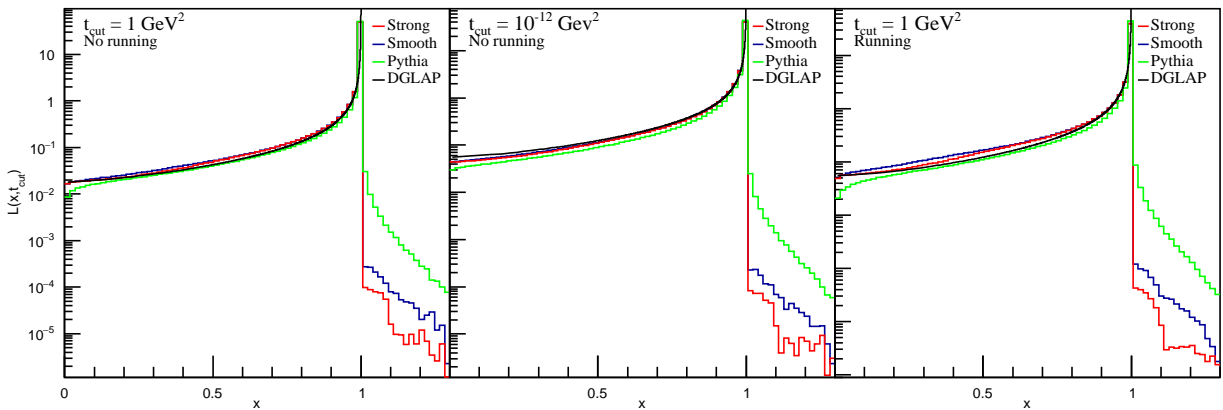}
\caption{Comparison between a numerical solution to the DGLAP equation, the sector shower approach and the default Pythia shower. In the left and middle pane, the coupling $\alpha$ is kept fixed at its value at the electron mass, while the cutoff scale is varied. On the right, $\alpha$ is allowed to run.}
\label{DGLAPcomparison}
\end{figure}

The result of the comparison is shown in figure \ref{DGLAPcomparison}. The hard scale $Q^2$ is set to the minimum of the invariant masses of all pairs of leptons for both the numerical DGLAP solution and the parton shower. This is the highest scale such that all sectors are able to radiate. The center of mass energy of the RAMBO event is set to $10^4 \, \mbox{GeV}$. In the left and middle panel, the shower cutoff $t_{\mbox{\tiny{cut}}}$ is set to $\Lambda_{\mbox{\tiny{QCD}}}^2 \approx 1 \, \mbox{GeV}^2$ and $t_{\mbox{\tiny{cut}}} = 10^{-12} \, \mbox{GeV}^2$, which is the default Pythia cutoff for photonic radiation. The coupling $\alpha$ is fixed to the default Pythia value at the electron mass $\alpha(m_e^2) = 0.00729735$. In the right panel, $\alpha$ is allowed to run from this value according to eq.~\eqref{running}. To enhance the effects of the running of $\alpha$, we set $n_f$ to $35$.

For both the Pythia shower and the sector approach, events appear with $x>1$. This is not possible in the analytical approach, but it is allowed in the shower since the $2 \rightarrow 3$ kinematics will sometimes raise the energy of a participating lepton. Outside this region, we observe strong agreement between both showers and the analytical approach when $\alpha$ is kept fixed. The agreement worsens when $\alpha$ is allowed to run. This is caused by the difference in scales used as argument for $\alpha$ in all three approaches. However, neither of the showers performs significantly better than the other. We note that, as no equivalent to the CWM scheme \cite{CMWscheme} exists for QED, there is no a priori preference for any scale, which is reflected in this result. 

\subsection{Effects of coherence}
In the currently available parton shower approaches to photon radiation such as those of PYTHIA or PHOTOS, not all eikonal factors are included. Instead, independent dipoles are constructed such that every radiating particle is assigned a single kinematic partner, usually of opposite sign to allow for a simple probabilistic interpretation. The correct collinear limits can be achieved through the normal $1 \rightarrow 2$ or $2 \rightarrow 3$ shower schemes, as well as the eikonal factor for these particle pairs only. However, all other eikonal factors are not included. Here, we compare the sector approach with such an incoherent strategy in the antenna formalism.

\begin{figure}[!htb]
\centering
\captionsetup{margin=.1\linewidth}
\includegraphics[scale=0.8]{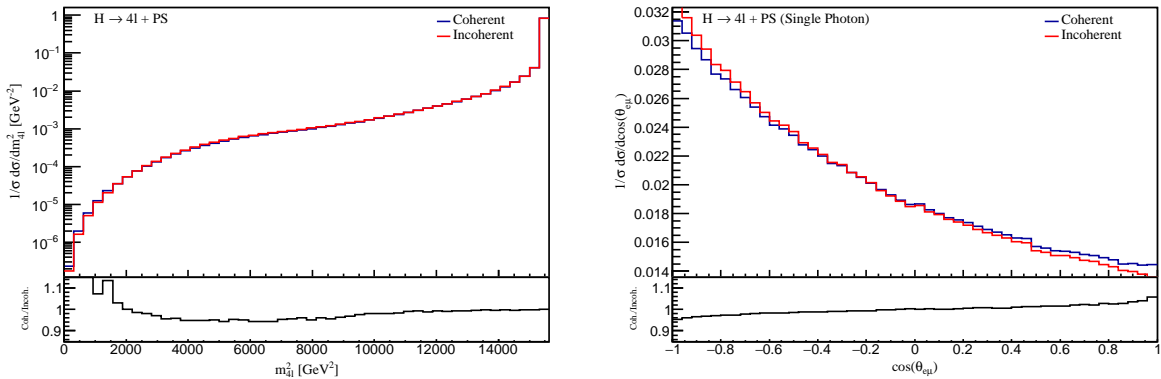}
\caption{Invariant mass distribution of the leptons (left) and the angle between the $e^-$ and the $\mu^-$ for single photon events (right) after application of the incoherent shower and the coherent sector approach for a Higgs mass of $125 \, \mbox{GeV}$. The parameters $\alpha$ and $t_{\mbox{\tiny{cut}}}$ are set to the default Pythia values.}
\label{coherence}
\end{figure}

Since the methods are equivalent for just two final state radiators, we consider the LHC-relevant decay process $H \rightarrow ZZ \rightarrow e^- e^+ \mu^- \mu^+$. The left-hand plot in figure \ref{coherence} shows the invariant mass distribution of the leptons after application of the showers. The difference between the algorithms is minor, only appearing at the very end of the mass spectrum. The coherent branching kernel can vary from being a factor of 2 larger than the incoherent branching kernel, to completely vanishing due to destructive interference. In case of the invariant lepton mass, the differences are largely averaged out. On the right-hand side, the distribution for the angle between the electron and muon is shown for events with exactly one photon. This variable separates phase space points where the difference between the coherent and incoherent branching kernels are most pronounced. 

\subsection{Comparison with YFS simulation}
\begin{figure}[!htb]
\centering
\captionsetup{margin=.1\linewidth}
\includegraphics[scale=0.8]{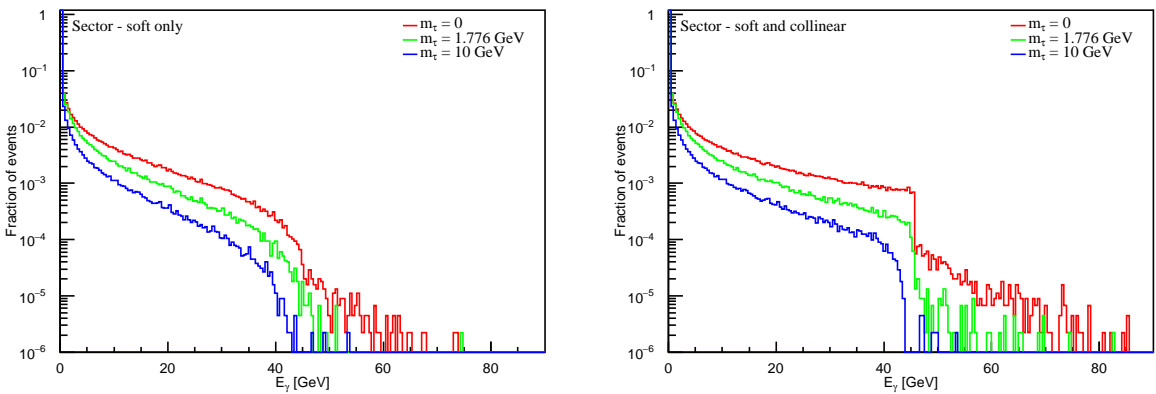}
\caption{Profile of the radiated energy in photons $E_{\gamma}$ off $Z \rightarrow 2\tau$ for several values of $m_{\tau}$.}
\label{YFS}
\end{figure}

In this section, we perform a brief comparison with the implementation of the YFS formalism as implemented by \cite{decays1,decays2}. The YFS formalism incorporates all soft logarithms, but collinear logarithms have to be included order-by-order, similar to matrix element corrections in parton showers. The sector approach includes both the soft and the collinear logarithms without resorting to corrections. To confirm that the soft behaviour of the sector approach is consistent with the YFS method, we display the photon radiation profiles for $Z \rightarrow 2\tau$ in figure \ref{YFS}. These radiation profiles are also shown in figure 1 in both \cite{decays1} and \cite{decays2} and we observe good agreement. In all cases $t_{\mbox{\tiny{cut}}}$ is set to $0.01 \, \mbox{GeV}^2$ and strong ordering is used. In the left-hand graph, the collinear single-pole terms of eq.~\eqref{aQED} are turned off, revealing their influence on the preference for hard photon production. The graphs drop off sharply at $E_{\gamma} = m_{Z}/2$ due to kinematic constraints. Higher values of $E_{\gamma}$ can only be reached more than one photon is emitted, which is rare in this decay. 

At the particle level, the YFS method can also be used to simulate photonic radiation off $W$ decay by treating emissions off the $W$ as initial state radiation. This is not yet possible in our approach, and we reserve this for later work in a full electroweak parton shower. In such a shower, it makes sense to treat $W$ and $Z$ decay as part of the shower similar to photon splitting. If these decays are just components of the shower, the $W$ is allowed to radiate before it eventually decays, and the decay product radiate afterwards.  

\subsection{Phase space discontinuities}

One concern with eq.~\eqref{aQED} and the sector shower approach is the presence of discontinuities in the radiative phase space on the boundaries between sectors. These discontinuities do not affect the formal accuracy of the shower since the collinear regions are far away from the boundary and in the soft region the fermion momenta are hardly changed. However, there may be an effect for high-scale photon emissions which is relevant for a potental implementation of matching and merging where the entire phase space has to be covered. 

To test for artifacts of these discontinuities, we compare the shower to events generated directly according to the branching kernel. The parton shower is run from the kinematic limit on a RAMBO-generated four charged lepton event with $E_{CM} = 10^4 \, \mbox{GeV}$ and $t_{\mbox{\tiny{cut}}} = 1 \, \mbox{GeV}^2$. The shower is terminated after a single emission, and only the events with an emission are kept. To remove the Sudakov suppression, a CKKW-L-like \cite{CKKW-L1,CKKW-L2} procedure is used where events are rejected with a probability that is generated using trial emissions from the scale of the actual emission. A directly generated event sample was compared with the unweighted parton shower sample, both with $\mathcal{O}(10^9)$ events, in the emission scale, the photon energy and the various leptonic invariant masses. The samples match up extremely well for all variables, giving no cause for concern for an implementation of matching or merging to matrix element calculations at a later stage.   

\subsection{Performance testing} \label{performanceTesting}
\begin{figure}[!htb]
\centering
\captionsetup{margin=.1\linewidth}
\includegraphics[scale=0.8]{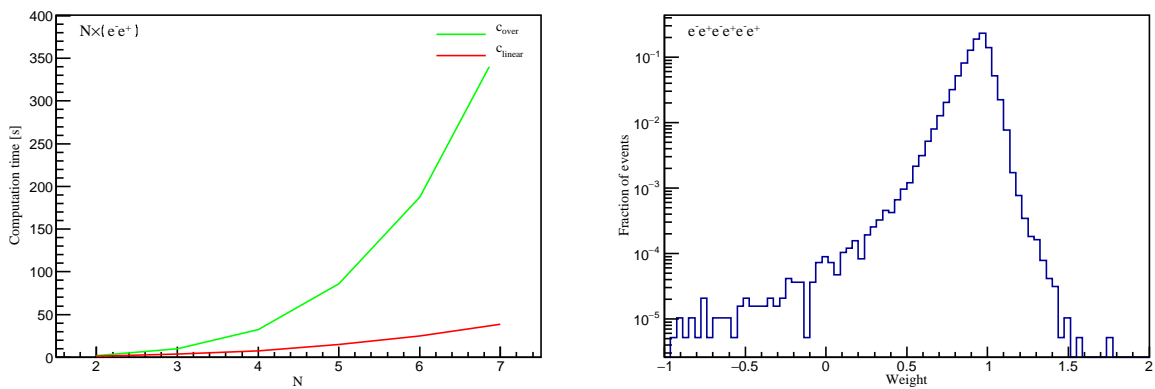}
\caption{Performance comparison of the regular veto algorithm using overestimate $c_{\mbox{\tiny{over}}}$ and algorithm \ref{SVAreweighting} using overestimate $c_{\mbox{\tiny{linear}}}$. On the left, the computation time to shower $10^4$ RAMBO-generated events of $N$ electron-positron pairs is compared. On the right, the distribution of weights that results from the application of the shower using algorithm \ref{SVAreweighting} on events with three electron-positron pairs is shown.}
\label{performance}
\end{figure}

We perform a brief performace comparison between the regular veto algorithm using the overestimate given by eq.~\eqref{simpleOverest} and algorithm \ref{SVAreweighting} using the overestimate give by eq.~\eqref{cLinear}. In figure \ref{performance} their relative performance for final states with an increasing number of charged leptons and a typical distribution of shower weights are shown. All events are produced with $E_{CM} = 10^4 \, \mbox{GeV}$ and the cutoff scale is set to $t_{\mbox{\tiny{cut}}} = 10^{-6} \, \mbox{GeV}^2$. The increase in performance is substantial and the weight distribution peaks strongly at one. Negative weights can occur when a trial scale is rejected in a region where the branching kernel is larger than the overestimate. The acceptance probability eq.~\eqref{sigmoid} is close to unity in these regions, so the probability for the appearance of negative weights is strongly suppressed. Note that the incomplete overestimate of eq.\eqref{cLinear} is by no means the only possible choice. If the weights are found to fluctuate too much, the overestimate can be raised at the cost of some performance. 

\section{Summary and Conclusion}\label{conclusionSection}
We described a formalism for coherent QED radiation in parton showers that is closely related to QCD antenna showers like ARIADNE and VINCIA. For photon radiation, all soft and collinear singularities are captured in a single branching kernel that is active over all of phase space. The phase space itself is divided into sectors such that branching can be regulated by an ordering parameter that remains similar to the standard antenna shower choice and the usual $2 \rightarrow 3$ kinematics can be used. A modified version of the Sudakov veto algorithm is presented to improve performance at the cost of introducing weighted events. For photon splitting, the methodology is much closer to the QCD analogon of gluon splitting with the exception of the presence of a color structure that can be used to dictate which spectator is used. A solution is provided by a generalized version of the so-called ARIADNE factor. 

For validation, we presented several comparisons with exact matrix element calculations, the DGLAP equation and the YFS method. When performing a full phase space scan, the shower approximation shows agreement with matrix elements at a similar level as the QCD counterpart in VINCIA. Good agreement is observed with both the DGLAP equation and the YFS radiation pattern. We also compared the sector approach with an incoherent shower similar to those implemented in PYTHIA and PHOTOS for a final state with four radiators at LHC energies. The differences are currently minor, but they should become increasingly relevant at future colliders as multiple relevant radiators appear in final states, especially once initial state radiation is also included.

In the near future, the QED shower formalism described in this paper will be implemented in the VINCIA parton shower. This implementation should also include initial state radiation, which is a relatively straightforward extension of the work presented in this paper. As a consequence, all relevant interference between initial and final state radiation will be included by construction. Initial state radiation and its interference with final state radiation has already been shown to be relevant for precision measurements at the LHC \cite{precision,precision2} and for future colliders \cite{future,future2}. In the future, helicity dependence and an extension to a full electroweak formalism will also be included.

\section*{Acknowledgement}
We would like to thank Peter Skands for communication and very helpful comments on the manuscript. This work was supported by the Netherlands Foundation for Fundamental Research of Matter (FOM) programme
entitled "Higgs as Probe and Portal".

\newpage
\appendix
\numberwithin{equation}{section}
\section{Antenna phase space factorization} \label{PSF}
Here we show how phase space can be factorized as indicated by eq.~\eqref{antennaPhaseSpace}. We first note that the 2-body phase space is

\begin{align} \label{2pps}
d\Phi_2 &= (2\pi)^{-2} \,d^4p_u \,d^4p_v \delta(p_u^2-m_u^2) \delta(p_v^2-m_v^2) \delta^4(P-p_u-p_v) \nonumber \\
&=\frac{1}{32m_{uv}^2 \pi^2} \lambda^{\frac{1}{2}}\left(m_{uv}^2, m_u^2, m_v^2\right)d\Omega_2
\end{align}
where $\lambda$ is the K{\"a}ll{\'e}n function. We now start from the $(n+1)$-body phase space where we explicitly factorize three momenta $p_a$, $p_b$ and $p_c$

\begin{align}
d\Phi_{n+1} &= (2\pi)^{4-3n} \frac{d^3p_a}{2E_a} \frac{d^3p_c}{2E_c} \delta((Q-p_a-p_c)^2-m_b^2) \prod_{i \neq a,b,c} d^4p_i \delta(p_i^2-m_i^2)
\intertext{where we denoted $Q = P - \sum_{i \neq a,b,c} p_i$. By a straightforward change of variables, this can be written as}
&= \frac{(2\pi)^{4-3n}}{32m_{abc}^2} \,ds_{ab} \,ds_{ac} \,ds_{bc} \,d\phi \,d\Omega_2 \delta\left(m_{abc}^2-s_{ab}-s_{ac} - s_{bc} - m_a^2 - m_b^2 - m_c^2\right) \nonumber \\
&\times \theta\left(G_{abc}>0\right) \prod_{i \neq a,b,c} d^4p_i \delta(p_i^2-m_i^2)
\end{align}
where $G_{ab} = s_{ab}s_{bc}s_{ac}-s_{ab}^2m_c^2 - s_{ac}^2m_{b}^2 - s_{bc}^2m_{a}^2 + 4m_{a}^2m_{b}^2m_{c}^2$ is the three-body Gram determinant. We can now replace the solid-angle integral by the 2-body phase space and absorb it into the remaining $(n-2)$-body piece to find the phase space factorization of eq.~\eqref{antennaPhaseSpace} where

\begin{align}
d\Phi^{ab}_{\mbox{\tiny{ant}}} = \frac{d\Phi_{3}}{d\Phi_2} &= \frac{1}{16 \pi^2}\lambda^{-\frac{1}{2}}(m_{uv}^2,m_u^2,m_v^2)\,ds_{ab} \,ds_{ac} \,ds_{bc} \frac{d\phi}{2\pi} \nonumber \\
&\times \delta\left(m_{abc}^2-s_{ab}-s_{ac} - s_{bc} - m_a^2 - m_b^2 - m_c^2\right) \theta\left(G_{abc}>0\right)
\end{align}
where the pre-branching particles are generally labelled with $u$ and $v$, but which are related to $a$ and $b$ in a way dependent on the branching process. 

\newpage
\bibliographystyle{JHEP}
\bibliography{ref}

\end{document}